\documentclass[aps,prx,onecolumn,superscriptaddress,floatfix,nofootinbib,10pt]{revtex4-2}

\usepackage[linesnumbered,lined,boxed,commentsnumbered,ruled,vlined]{algorithm2e}
\usepackage{algpseudocode}
\usepackage{amsfonts}
\usepackage{amsmath}
\usepackage{amssymb}
\usepackage{graphicx}
\usepackage{dcolumn}
\usepackage{bm,bbm}
\usepackage{tabularx}
\usepackage{epstopdf}
\usepackage{xcolor}
\usepackage{mathrsfs}
\usepackage{mathtools}
\usepackage{float}
\usepackage{verbatim}
\usepackage{comment}
\usepackage{ragged2e}
\usepackage{physics}
\usepackage{hyperref}
\usepackage{url}
\usepackage{cleveref}
\usepackage{multirow}
\usepackage{scalerel}
\usepackage{eurosym}
\usepackage{booktabs}

\hypersetup{colorlinks=true, citecolor=orange, urlcolor=blue, linkcolor=magenta}

\begin{document}

\title{Topology as information: network effects in corporate lending}

\author{Anna Pirogova}
\affiliation{IMT School for Advanced Studies, P.zza San Francesco 19, 55100 Lucca (Italy)}
\author{Anna Mancini}
\affiliation{Department of Physics and INFN, `Tor Vergata' University of Rome, Via della Ricerca Scientifica 1, 00133 Rome (Italy)}
\affiliation{`Enrico Fermi' Center for Study and Research (CREF), Via Panisperna 89A, 00184 Rome (Italy)}
\author{Tiziano Squartini}
\affiliation{IMT School for Advanced Studies, P.zza San Francesco 19, 55100 Lucca (Italy)}
\affiliation{Scuola Normale Superiore, P.zza dei Cavalieri 7, 56126 Pisa (Italy)}
\affiliation{Istituto Nazionale di Alta Matematica `Francesco Severi' (INdAM-GNAMPA), P.le Aldo Moro 5, 00185 Rome (Italy)}
\author{Giulio Cimini}
\affiliation{Department of Physics and INFN, `Tor Vergata' University of Rome, Via della Ricerca Scientifica 1, 00133 Rome (Italy)}
\affiliation{`Enrico Fermi' Center for Study and Research (CREF), Via Panisperna 89A, 00184 Rome (Italy)}

\begin{abstract}
A central challenge in financial economics is understanding how credit networks form under informational noise. We introduce the concept of \emph{topological capital}, arguing that banks increasingly rely on \emph{topological certification}, interpreting a borrower's connectivity as a primary proxy for creditworthiness. Using a novel dataset of bank-firm relationships manually extracted from Italian financial statements, we implement a multi-stage empirical framework, benchmarking empirical patterns against a maximum-entropy benchmark, to separate the determinants of credit access from those of loan volumes. Our results indicate that network topology systematically outperforms traditional fundamentals. In the link-formation stage, connectivity breeds further connectivity through an amplified preferential attachment mechanism. In the loan-sizing stage, network strength absorbs the explanatory power of balance-sheet metrics, documenting a profound \emph{network substitution effect} where topological signals effectively replace physical collateral across all corporate segments. For SMEs, we identify a critical signal divergence: reported debt acts as a risk signal, while network footprint serves as market validation. Furthermore, we reveal a \emph{diversification paradox}: while firms fragment debt to avoid hold-up risks, over-diversification leads to a \emph{complexity penalty} that stagnates credit depth and inflates systemic Loss Given Default. Ultimately, our findings signal the \emph{twilight of the balance sheet} as the primary anchor of corporate lending, calling for a shift toward topological macro-prudential supervision to manage vulnerabilities invisible to traditional bilateral indicators.
\end{abstract}

\maketitle

\section{Introduction}

Traditional corporate finance theory posits that banks act as delegated monitors~\cite{Diamond:1984aa}, deciding whether to lend to a firm by analysing hard borrower fundamentals, such as profitability (ROA), leverage, and collateral availability (tangibility). However, this view treats lending decisions as isolated bilateral contracts. Here, we argue that the credit market functions as an integrated information-processing system, where the existence of a link itself conveys information to other market participants, thus influencing systemic risk and stability~\cite{Battiston:2012aa,Puliga:2014aa,Bardoscia:2021aa,Tabachova:2024aa,Anand:2015aa,Piccardi:2011aa,Mistrulli:2011aa}.

The theoretical underpinning of our argument lies in the pervasive nature of information asymmetry in corporate lending, in particular for Small and Medium Enterprises (SMEs)~\cite{Berger:2006aa}. As demonstrated by Stiglitz and Weiss~\cite{Stiglitz:1981aa}, in markets where borrowers' risk types are opaque, interest rates alone cannot function as a clearing mechanism due to adverse selection; higher rates may simply attract riskier borrowers or incentivize safer borrowers to take excessive risks (moral hazard). Consequently, banks engage in credit rationing, limiting supply rather than raising prices. In this context of imperfect information, the cost of screening and monitoring becomes a critical determinant of market structure~\cite{Boot:2000aa}. In this landscape, the phenomenon of multiple banking - ubiquitous in Italy - emerges as a strategic equilibrium: while firms maintain multiple relationships to avoid information monopolies and the `hold-up' problem associated with single-bank lending~\cite{Sharpe:1990aa,Rajan:1992aa}, this fragmentation paradoxically increases the complexity of evaluating creditworthiness, forcing lenders to look for external signals.

We posit that banks mitigate these high screening costs by relying on topological certification. Specifically, the topology of the corporate lending network serves as a mechanism for signal transmission: when direct screening is costly or financial statements are opaque, banks may interpret the lending decisions of peers as revealed information regarding a borrower's quality - a phenomenon akin to the observational learning described by Banerjee~\cite{Banerjee:1992aa}. This creates a strong incentive for `screening free-riding': rather than incurring the full cost of independent assessment, secondary lenders may rely on the certification provided by the lending decisions of incumbents, effectively treating the existence of a link as a hard proof of solvency. If a firm has already secured loans from multiple institutions, a new lender may infer that the firm has successfully passed the screening processes of those incumbents: this reliance on the `wisdom of the crowd' effectively allows banks to leverage on the system's collective monitoring efforts~\cite{Carletti:2007aa}.

This hypothesis suggests a mechanism of information cascades: banks utilize the network topology as a primary proxy for creditworthiness, potentially overriding fundamental financial metrics. In such a system, credit rationing is driven by a preferential attachment dynamics~\cite{Barabasi:1999aa}, where connectivity breeds further connectivity. More importantly, we hypothesize a `network substitution effect' where topological validation acts as a form of \emph{reputational collateral} that effectively replaces the need for physical fixed assets. Building on this, we introduce the concept of \emph{topological capital}, i.e. the idea that a firm's position within the credit network is not a mere outcome of its financial health, rather a strategic asset itself. In an era of increasing informational noise, a firm's connectivity serves as a `distributed ledger' of trust that can be more valuable - and more liquid - than the physical assets on its balance sheet.

The Italian context offers an ideal laboratory for testing these competing hypotheses, due to the rich structure of its banking layers~\cite{Bargigli:2015aa}, the landscape of firms dominated by SMEs and the existence of \emph{Centrale Rischi} (Central Credit Register). Centrale Rischi is a public credit information system managed by Bank of Italy, which helps banks to assess the creditworthiness of loan applicants by providing information on their \emph{Accordato di Sistema} (aggregate system exposure). Although these data remain confidential to researchers, we construct a unique dataset in a bottom-up fashion, by manually extracting lending relationships from the \emph{Nota Integrativa} (notes to financial statements) of Italian firms.

Our analysis on these extracted bank-firm relationships, based on a multi-stage empirical framework, provides a fourfold contribution. First, we provide robust evidence that network variables consistently outperform traditional financial metrics in predicting both credit access and loan volumes. Second, we document a profound `network substitution effect': once topological signals are accounted for, the explanatory power of asset tangibility vanishes across corporate segments, suggesting that network validation acts as reputational collateral in the modern credit market. Third, we reveal a critical signal divergence for SMEs, where reported balance sheet debt is interpreted as a risk signal (negative coefficient) while observed network strength acts as a validation signal (positive coefficient). Finally, by benchmarking our results against a maximum-entropy benchmark~\cite{Squartini:2017aa}, we identify a structural complexity penalty for large groups, where over-diversification appears as linked to a potential stagnation in credit depth. Taken together, our findings call for a shift towards a \emph{topological supervision} to manage the systemic risk that is inherent to a network-driven credit market.

\section{Data description}

\subsection{Data sources and sample construction}

To investigate the determinants of lending relationships, we constructed a dataset of bank-firm linkages by manually extracting lending relationships from the \emph{Nota Integrativa} (notes to financial statements) of Italian firms for the fiscal year $2022$. The analysis is conducted on two distinct samples to ensure robustness across different corporate structures:

\begin{itemize}
\item Sample A (the \emph{consolidated} one) focuses on large corporate groups: we selected $113$ firms identified as ultimate owners of their respective groups, based on the availability of detailed bank debt breakdowns in the financial statements. Firms were selected in descending order of total assets, with a debt to banks above \euro\:$1.000$. This selection process resulted in a sample characterized by a minimum total assets threshold of \euro\:$300.000.000$. Financial data refers to their consolidated balance sheets, sourced from Bureau van Dijk's AIDA;
\item Sample B (the \emph{unconsolidated} one) includes a broader set of $175$ firms. The selection followed a similar descending order of total assets, resulting in a minimum total assets threshold of \euro\:$75.000.000$. Financial data refers to their unconsolidated balance sheets, sourced from Bureau van Dijk's AIDA.
\end{itemize}

We, then, built the banking layer by identifying the lending institutions for both samples ($61$ and $64$ banks for the consolidated and unconsolidated sample, respectively). Financial data refers to unconsolidated balance sheets, for the fiscal year $2022$: basic data was sourced from Bureau van Dijk's AIDA, while granular data on corporate loans was manually extracted from the balance sheets files available on Bureau van Dijk's BankFocus, to ensure a precise distinction between corporate and retail lending portfolios.

Each sample is finally represented as a bipartite network $\mathbf{A}\equiv\{a_{ij}\}$ of $N_F$ firms and $N_B$ banks (i.e. a rectangular matrix with $N_F$ rows and $N_B$ columns): the presence of a loan contract between firm $i$ and bank $j$ corresponds to a link ($a_{ij}=1$) with weight $w_{ij}>0$ (the amount of the loan), whereas $w_{ij}=a_{ij}=0$ if $i$ and $j$ are disconnected.

\subsection{Definition of variables}

We categorize our determinants into two competing frameworks. The first one is defined by network variables: the network strength of firm $i$, 
$s_i^{net}=\sum_{j=1}^{N_B}w_{ij}$, is defined as the sum of all financing links incident to it and proxies the total credit granted by the system (\emph{Accordato Operativo}), while the degree of firm $i$, $k_i=\sum_{j=1}^{N_B}a_{ij}$, is defined as the number of distinct banking relationships maintained by the firm and proxies the number of reporting banks (\emph{Numero di Segnalanti}). Analogously, the network strength of bank $j$, $t_j^{net}=\sum_{i=1}^{N_F}w_{ij}$, proxies the total credit provided to the system by the bank, while the degree of bank $j$, $h_j=\sum_{i=1}^{N_F}a_{ij}$, proxies the number of loan contracts issued by that bank. 

The second one is defined by fundamental variables. For firms, we employ standard corporate balance sheet figures and finance ratios sourced from Bureau van Dijk's AIDA: debt to banks reported in liabilities (which we refer to as balance sheet strength, indicated as $s_i^{bal}$ for firm $i$); asset tangibility, defined as the ratio of tangible fixed assets to total assets and serving as a proxy for collateral availability; profitability (ROA), defined as net income divided by total assets; risk, measured by the leverage (total liabilities over total assets). For banks, we control the supply factors using the corporate loans reported in the asset side of the balance sheet (which again we refer to as balance sheet strength, indicated as $t_j^{bal}$, for bank $j$); total assets; ROA; leverage.

We remark that our bottom-up data collection process necessarily focuses on a significant, but non-exhaustive, subset of the Italian corporate landscape. Hence, we expect that the network strengths of banks ($t^{net}_j$, $\forall\:j$) will be systematically lower than the total corporate loans reported on their official balance sheets ($t^{bal}_j$, $\forall\:j$), since the latter include loans to firms outside our samples. Nevertheless, we anticipate these two measures to be highly correlated, confirming that our firm selection effectively captures the core of the corporate credit market where topological signals are generated.

Regarding firms, network strength ($s^{net}_i$, $\forall\:i$) and balance sheet debt ($s^{bal}_i$, $\forall\:i$) may diverge due to two primary factors. First, our manual extraction from the \textit{Nota Integrativa} captures granular off-balance-sheet items, such as guarantees and abnormal financing arrangements, that are explicitly linked to specific lenders but are typically excluded from the official `Debt to Banks' line item in standard financial statements. Second, disclosure practices in Italy often lead firms to explicitly name only institutional or major lenders in their financial notes, while leaving other secondary relationships unnamed or reported in aggregate. Consequently, the sum of individual financing volumes identified through our data reconstruction ($s^{net}_i$, $\forall\:i$) is unlikely to perfectly match the aggregate balance sheet figure ($s^{bal}_i$, $\forall\:i$). In this sense, our network metrics should be interpreted as a granular representation of a firm's primary credit ecosystem, rather than a strict accounting identity.

Given these observations, to increase the reliability of the manually-constructed network, we applied a consistency band filter. After calculating the firm-specific ratio between the (constructed) network strength $s_i^{net}$ and the (reported) balance sheet strength $s_i^{bal}$, we retain only those whose ratio falls within the range $[10^{-3},10^3]$. This means that we filter out observations where either the manual data collection process failed to capture a representative portion of the firm's financial reality (including firms with less than $0.1\%$ of their reported debt would imply treating `missing data' as `low connectivity', thus generating spurious results), or an extremely inconsistent \emph{Nota Integrativa} (firms that report individual exposures larger than $1000\%$ of their total declared bank debt).

\begin{table}[b!]
\centering
\begin{tabular}{lccc}
\hline
\hline
& & \textbf{Consolidated} & \textbf{Unconsolidated} \\
\hline
\hline
Number of firms & $N_F$ & 113 & 175 \\
Number of banks & $N_B$ & 61 & 64 \\
Link density & $L_{obs}/(N_F N_B)$ & 0.07 & 0.05 \\
Average firm degree & $\overline{k}=L_{obs}/N_F$ & 4.34 & 3.37 \\
Average bank degree & $\overline{h}=L_{obs}/N_B$ & 7.97 & 9.14 \\
Firms' degree CV & $\text{CV}_F=\sigma_{k}/\overline{k}$ & 0.75 & 0.80 \\
Banks' degree CV & $\text{CV}_B=\sigma_{h}/\overline{h}$ & 1.83 & 1.96 \\
\hline
\hline
\end{tabular}
\caption{Summary network statistics for the consolidated and unconsolidated sample.}
\label{tab:network_stats}
\end{table}

\subsection{Descriptive statistics}

\begin{figure*}[t!]
\centering
\large {\bf Consolidated sample}
\includegraphics[width=\textwidth]{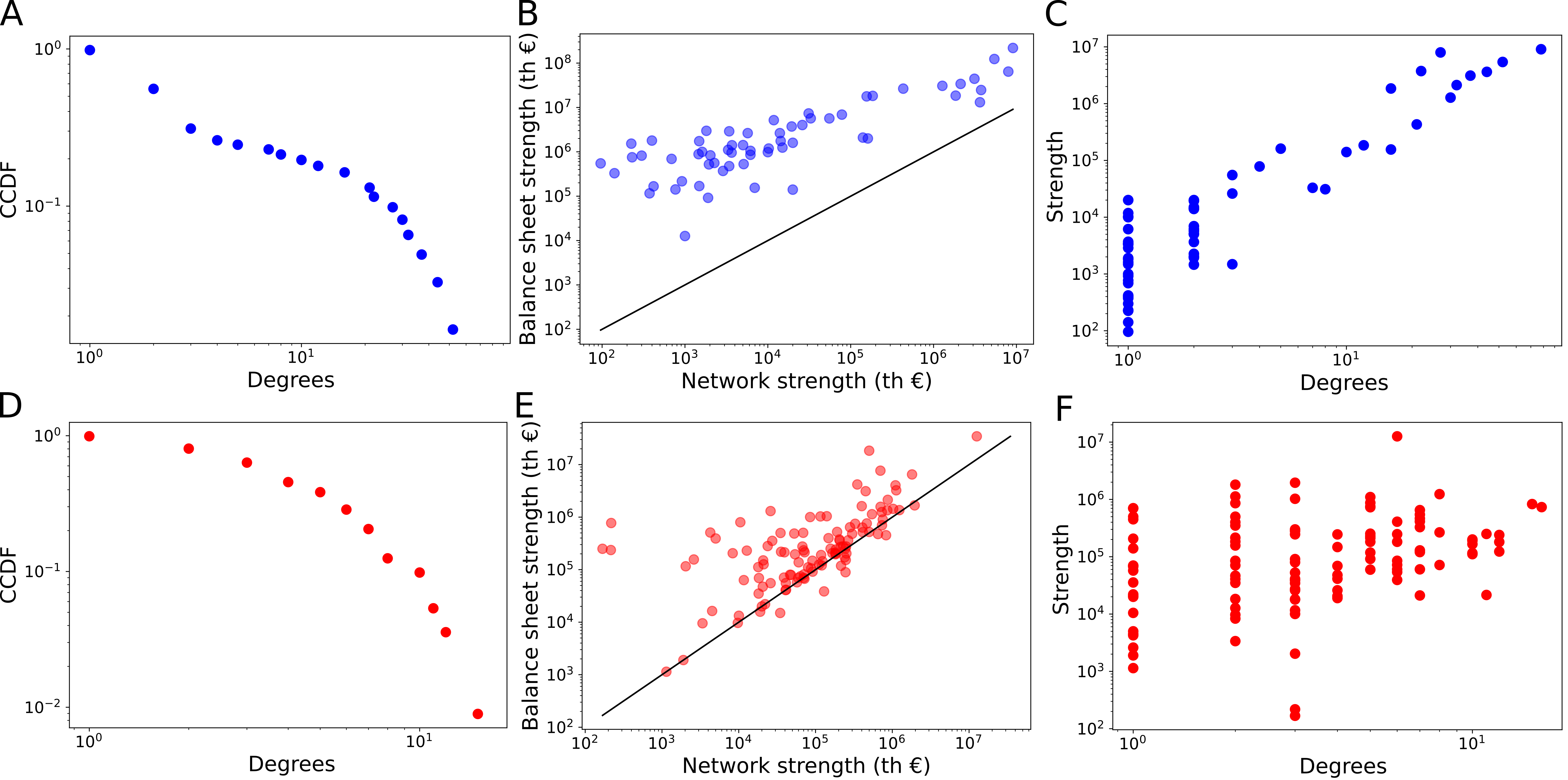}\\
\smallskip
\large {\bf Unconsolidated sample}
\includegraphics[width=\textwidth]{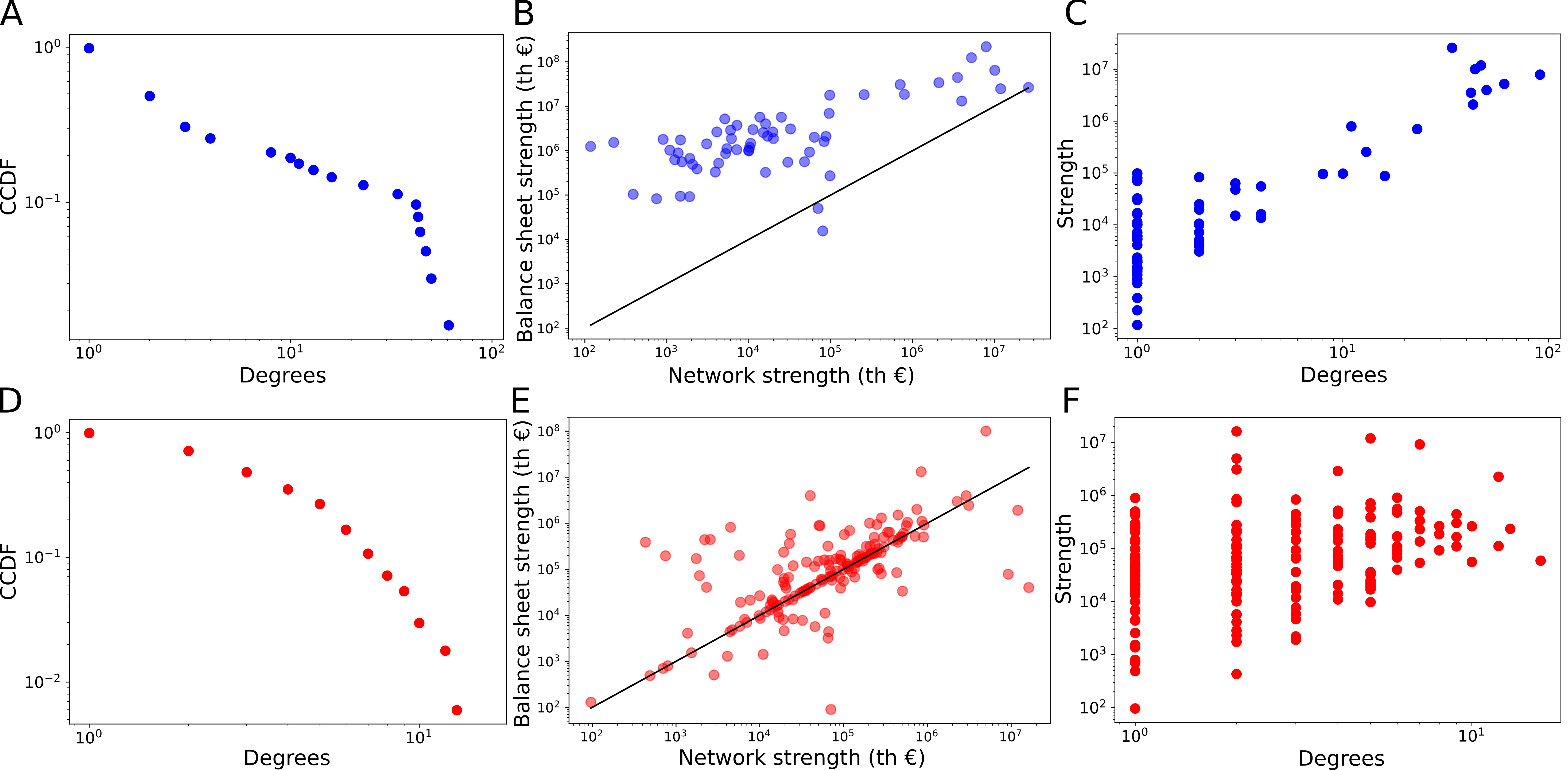}
\caption{\textbf{Empirical network statistics for the consolidated (top) and unconsolidated (bottom) samples.} Panels (A,B,C) refer to bank variables while panels (D,E,F) to firm variables. (A,D) Complementary cumulative distribution function of degrees, showing that bank connectivity is more heterogeneous than firm connectivity. (B,E) Balance sheet strength versus network strength (the solid line marks the identity): for banks, the offset is due to our data collection procedure, not covering all corporate loans of banks. (C,F) Network strength versus degree: size and number of relationships correlate well for banks but only modestly for firms, for which the variability is large for poorly-connected nodes.}
\label{fig1}
\end{figure*}

Figure~\ref{fig1} and Table~\ref{tab:network_stats} report network statistics for the consolidated and unconsolidated sample. For what concerns banks, connectivity is very heterogeneous due to the presence of a few large banks acting as hubs that lend to several firms (Panel A). The balance sheet strength is well-correlated with, but systematically higher than, the network strength, due to our data collection procedure that focuses on firms loans and may, thus, not cover all bank loans (Panel B). Network strength grows more than proportionally with bank degree, implying that large banks with many client firms have much higher loan-to-deposit ratios than small banks with a few clients (Panel C).

Concerning firms, although the degree distribution is not fat-tailed, they are typically financed by multiple banks (Panel D). Balance sheet and network strengths mostly coincide, the outliers having been filtered out as described above (Panel E). Network strength and number of relationships are not highly correlated, with a larger variability observed for low-degree firms (Panel F). 

Overall, the patterns we observe are consistent with those reported by De Masi and Gallegati~\cite{De-Masi:2012aa}.

\section{Maximum-entropy counterfactual}

We, now, define a benchmark of our bank-firm network by using the maximum-entropy prescription, proposed in~\cite{Squartini:2017aa} for capital markets and described in detail in Appendix~\hyperlink{AppA}{A}. In a nutshell, the model defines an ensemble of networks, where a single network instance is constructed with the following two-step procedure~\cite{Cimini:2015aa}:\\

\emph{i)} first, we establish a link between firm $i$ and bank $j$ with probability

\begin{equation}
p_{ij}=\frac{zs_it_j}{1+zs_it_j},
\end{equation}
where $z$ is a constant set to preserve the link density of the empirical network, while $s_i$ and $t_j$ are proxies of the firm's and bank's size, respectively. Therefore, the expected degrees read $\langle k_i\rangle=\sum_{j=1}^{N_B}p_{ij}$ and $\langle h_j\rangle=\sum_{i=1}^{N_F}p_{ij}$, i.e. solely driven by the firm's and bank's size - the so-called \emph{fitness ansatz}~\cite{Caldarelli:2002aa};\\

\emph{ii)} second, each link established in the first step is assigned the weight

\begin{equation}
w_{ij|_{a_{ij}=1}}=\frac{s_it_j}{Wp_{ij}}
\end{equation}
in a conditional fashion, the normalization constant reading $W=\sqrt{ST}$ where $S=\sum_{i=1}^{N_F}s_i$ and $T=\sum_{j=1}^{N_B} t_j$ are the total size of firms and banks, respectively: as a consequence, the expected link weight reads $\langle w_{ij}\rangle=s_i t_j/W$, while the expected strengths read $\langle s_i\rangle=\sum_{j=1}^{N_B}\langle w_{ij}\rangle=s_i\,T/W$ and $\langle t_j\rangle=\sum_{i=1}^{N_F}\langle w_{ij}\rangle = t_j\,S/W$.

In what follows, we employ two different variants of our maximum-entropy counterfactual: a network-driven version, where sizes are proxied by the network strengths, i.e. the actual credit volumes observed in the network ($s_i=s^{net}_i$, $\forall\:i$ and $t_j=t^{net}_j$, $\forall\:j$) and a balance-driven version, where sizes are proxied by the balance sheet strengths, i.e. the \emph{Debt to Banks} for firms and the \emph{Corporate Loans} for banks ($s_i=s^{bal}_i$, $\forall\:i$ and $t_j=t^{bal}_j$, $\forall\:j$). Notice that the network-driven variant ensures that $S=T$, i.e. that the total sizes of firms and banks coincide, hence the expected value of the strengths coincide with their empirical value: $\langle s_i\rangle=s^{net}_i$, $\forall\:i$ and $\langle t_j\rangle=t_j^{net}$, $\forall\:j$. In general, this is not the case for the balance sheet-driven variant of the model.

\begin{figure*}[t!]
\centering
\includegraphics[width=\textwidth]{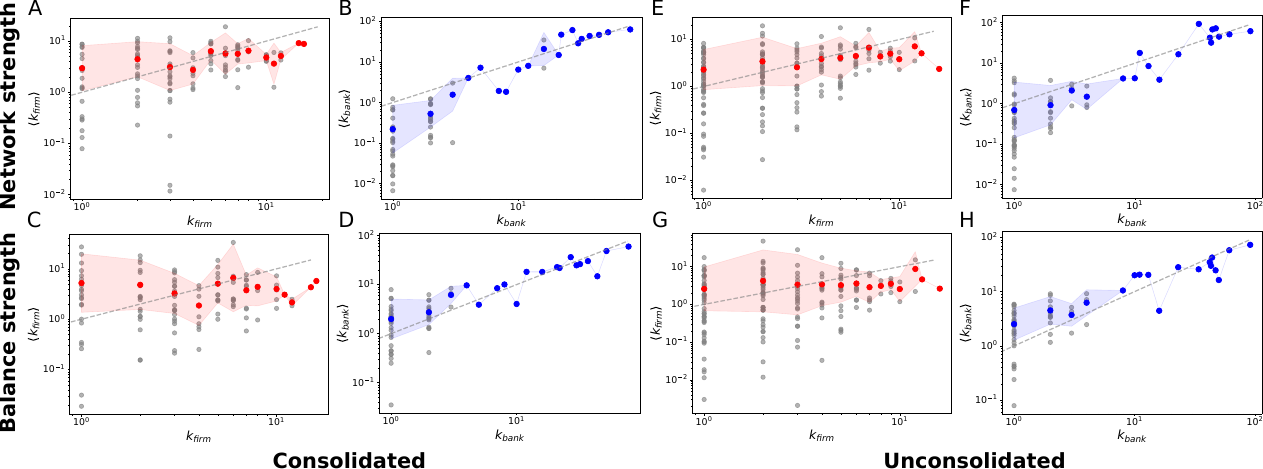}
\caption{\textbf{Empirical VS expected degrees.} Panels in the top (bottom) row refer to the network-driven (balance-driven) version of the model, while panels on the left (right) refer to the consolidated (unconsolidated) sample. The comparison is for firms' degrees (Panels A,C,E,G) and banks' degrees (Panels B,D,F,H). Benchmark predictions are obtained as averages over an ensemble of $10^4$ configurations. Gray dots show the raw data, coloured dots and shaded area show the mean and standard deviation of the model degrees over each real degree bin. The dashed line marks the identity.}
\label{fig2:cons-uncons}
\end{figure*}

Figure~\ref{fig2:cons-uncons} shows the comparison between empirical and expected values of the degrees, across both samples. Let us, first, focus on the network-driven model: for what concerns banks, empirical and predicted values are well-aligned, the Spearman (Pearson) correlation coefficient amounting to $\rho\approx0.77$ ($0.94$) on the consolidated sample and to $\rho\approx0.73$ ($0.93$) on the unconsolidated sample; for what concerns firms, however, empirical and predicted values are not so well-correlated, the Spearman (Pearson) correlation coefficient amounting to $\rho\approx0.08$ ($-0.05$) on the consolidated sample and to $\rho\approx0.19$ ($0.04$) on the unconsolidated sample. A similar outcome is observed for the balance-driven version of the model.

The results above suggest that while the number of a bank's clients can be regarded as a function of the bank's lending capacity, a firm's connectivity can be hardly inferred by solely looking at its balance sheets data.

\section{Econometric analysis}

In order to test the determinants of credit allocation, we employ a two-step regression analysis.

\subsection{Step $\#1$: modelling credit access (link formation)}

First, we estimate the probability that a link exist by employing a logistic regression. In formulas,
\begin{align}
P_{ij}(a_{ij}=1|\mathbf{X}^F_i,\mathbf{X}^B_j)&=\Lambda(\alpha+\mathbf{X}^F_i\bm{\beta}_i+\mathbf{X}^B_j\bm{\gamma}_j)=\frac{1}{1+e^{-(\alpha+\mathbf{X}^F_i\bm{\beta}_i+\mathbf{X}^B_j\bm{\gamma}_j)}},
\label{logit}
\end{align}
where $\Lambda(\cdot)$ is the logistic cumulative distribution function, $\bm{\beta}_i$ and $\bm{\gamma}_j$ respectively represent the vector of firm-specific and bank-specific parameters, and $\alpha$ is the intercept. Parameters are numerically estimated by maximizing the log-likelihood function

\begin{align}
\mathcal{L}(\mathbf{X}^F_i,\mathbf{X}^B_j)&=\sum_{i=1}^{N_F}\sum_{j=1}^{N_B}\left[a_{ij}\ln P_{ij}\left(a_{ij}=1|\mathbf{X}^F_i,\mathbf{X}^B_j\right)+(1-a_{ij})\ln P_{ij}\left(a_{ij}=0|\mathbf{X}^F_i,\mathbf{X}^B_j\right)\right],
\label{logit_like}
\end{align}
where $P_{ij}\left(a_{ij}=0|\mathbf{X}^F_i,\mathbf{X}^B_j\right)=1-P_{ij}\left(a_{ij}=1|\mathbf{X}^F_i,\mathbf{X}^B_j\right)$. In what follows, we perform three distinct regressions to separate the effects imputable to balance sheet fundamentals from those imputable to the network topology.\\

\noindent\emph{Model 1: Gravity Model.} It tests the explanatory power of traditional financial metrics while excluding any topological information. This specification boils down to the standard gravity model~\cite{Tinbergen:1963aa}, where the interaction probability is determined by the masses of the borrower and the lender:
\begin{align*}
\mathbf{X}^F_i&=\left\{\ln s_i^{bal},\mathbf{F}_i\right\}\\
\mathbf{X}^B_j&=\left\{\ln t_j^{bal},\mathbf{G}_j\right\}
\end{align*}
where $\mathbf{F}_i=\{\ln(\text{Assets}_i),\text{Lev}_i,\text{ROA}_i,\text{Tang}_i\}$ and $\mathbf{G}_j=\{\ln(\text{Assets}_j),\text{Lev}_j,\text{ROA}_j\}$ represent the vector of firm-specific and bank-specific controls, respectively.\\

\noindent\emph{Model 2: Network Model.} It tests the explanatory power of network variables while excluding any traditional financial metrics:

\begin{align*}
\mathbf{X}^F_i&=\left\{\ln s_i^{net},\ln k_i\right\}\\
\mathbf{X}^B_j&=\left\{\ln t_j^{net},\ln h_j\right\}
\end{align*}
with obvious meaning of the symbols.\\

\noindent\emph{Model 3: Full Model.} It includes all variables to test which factors survive when controlling for the others:

\begin{align*}
\mathbf{X}^F_i&=\left\{\ln s_i^{bal},\mathbf{F}_i,\ln s_i^{net},\ln k_i\right\}\\
\mathbf{X}^B_j&=\left\{\ln t_j^{bal},\mathbf{G}_j,\ln t_j^{net},\ln h_j\right\}
\end{align*}

To avoid mechanical endogeneity where a node's topological metrics naturally include the relationship being predicted, we follow Herman~\cite{Herman:2022aa} and construct `rest-of-the-world' variables for both degrees and strengths: specifically, we calculate the degrees as $k_{i,-j}=k_i-a_{ij}$ and $h_{j,-i}=h_j-a_{ij}$ and the network strengths as $s^{net}_{i,-j}=s^{net}_i-w_{ij}$ and $t^{net}_{j,-i}=t^{net}_j-w_{ij}$. In words, existing links (i.e. $a_{ij}=1$) and their volumes are subtracted from the predictors; for non-existing links (i.e. $a_{ij}=0$), instead, the variables remain unchanged: this ensures that the regressors capture a firm's attractiveness to other lenders in the system independently from the specific bilateral tie to be predicted.

\subsection{Step $\#2$: modelling credit volume (loan sizing)}

Conditional on the existence of a link ($a_{ij}=1$), we estimate the loan size by employing an ordinary least squares (OLS) regression. In formulas,

\begin{equation}
\mathbb{E}\left[\ln w_{ij}|\mathbf{X}^F_i,\mathbf{X}^B_j, a_{ij}=1\right]=\alpha+\mathbf{X}^F_i\bm{\beta}_i+\mathbf{X}^B_j\bm{\gamma}_j+\epsilon_{ij}
\label{OLS}
\end{equation}
where $w_{ij}$ denotes the size of the loan between firm $i$ and bank $j$, $\bm{\beta}_i$ and $\bm{\gamma}_j$ respectively represent the vector of firm-specific and bank-specific parameters, $\alpha$ is a constant and $\epsilon_{ij}$ is the residual term. Parameters are numerically estimated by minimizing the sum of squared residuals

\begin{align}
\text{RSS}(\mathbf{X}^F_i,\mathbf{X}^B_j)&=\sum_{i=1}^{N_F}\sum_{j=1}^{N_B}\left[\ln w_{ij}-\left(\alpha+\mathbf{X}^F_i\bm{\beta}_i+\mathbf{X}^B_j\bm{\gamma}_j\right)\right]^2.
\label{ols}
\end{align}

Since this second step focuses on the subset of existing links, the mechanical endogeneity issue applies to both degrees and strengths; therefore, we apply a two-fold correction: \emph{i)} for what concerns the degrees, we subtract the existing tie, thus obtaining $k_i-1$ and $h_j-1$; \emph{ii)} for what concerns the strengths, we subtract the specific loan amount from network and balance-sheet measures for both firms and banks, thus obtaining $s^{net}_{i,-j}=s^{net}_i-w_{ij}$, $t^{net}_{j,-i}=t^{net}_j-w_{ij}$, $s^{bal}_{i,-j}=s^{bal}_i-w_{ij}$ and $t^{bal}_{j,-i}=t^{bal}_j-w_{ij}$. This ensures that our regressors capture a firm's diversification and a bank's exposure to the rest of the system independently from the specific bilateral contract being estimated.

\begin{table*}[t!]
\centering
\renewcommand{\arraystretch}{0.8} % Riduce lo spazio tra le righe per far rientrare la tabella
\setlength{\tabcolsep}{3pt}       % Riduce lo spazio tra le colonne
\resizebox{\textwidth}{!}{        %
\begin{tabular*}{\textwidth}{@{\extracolsep{\fill}}lccccc}
\hline
\hline
\multicolumn{6}{c}{\bf Panel A: Consolidated Sample} \\
\hline
\hline
& {\bf Model 1} & \multicolumn{2}{c}{\bf Model 2} & \multicolumn{2}{c}{\bf Model 3} \\
\hline
Variable & (Gravity) & A (w/ degree) & B (w/o degree) & A (w/ degree) & B (w/o degree) \\
\hline
Intercept & -17.58*** (0.70) & -5.97*** (0.40) & -7.55*** (0.35) & -0.63 (1.07) & -1.81** (0.71) \\
\addlinespace
$\ln s^{net}$ & -- & \begin{tabular}[t]{@{}c@{}} -0.14*** (0.03) \\ \relax [-0.0048] \end{tabular} & \begin{tabular}[t]{@{}c@{}} 0.00 (0.02) \\ \relax [0.0002] \end{tabular} & \begin{tabular}[t]{@{}c@{}} -0.17*** (0.04) \\ \relax [-0.0057] \end{tabular} & \begin{tabular}[t]{@{}c@{}} -0.01 (0.03) \\ \relax [-0.0003] \end{tabular} \\
$\ln k$ & -- & \begin{tabular}[t]{@{}c@{}} 1.51*** (0.11) \\ \relax [0.0516] \end{tabular} & -- & \begin{tabular}[t]{@{}c@{}} \textbf{1.55*** (0.12)} \\ \relax \ [0.0539] \end{tabular} & -- \\
$is\_exclusive$ & -- & \begin{tabular}[t]{@{}c@{}} 0.53 (0.60) \\ \relax [0.0181] \end{tabular} & -- & \begin{tabular}[t]{@{}c@{}} 0.53 (0.64) \\ \relax [0.0186] \end{tabular} & -- \\
$\ln s^{bal}$ & \begin{tabular}[t]{@{}c@{}} -0.01 (0.03) \\ \relax [-0.0004] \end{tabular} & -- & -- & \begin{tabular}[t]{@{}c@{}} -0.14*** (0.04) \\ \relax [-0.0049] \end{tabular} & \begin{tabular}[t]{@{}c@{}} -0.24*** (0.04) \\ \relax [-0.0102] \end{tabular} \\
$\ln\text{Assets}_{firm}$ & \begin{tabular}[t]{@{}c@{}} -0.02 (0.03) \\ \relax [-0.0008] \end{tabular} & -- & -- & \begin{tabular}[t]{@{}c@{}} 0.01 (0.03) \\ \relax [0.0002] \end{tabular} & \begin{tabular}[t]{@{}c@{}} -0.05** (0.02) \\ \relax [-0.0022] \end{tabular} \\
$\text{Lev}_{firm}$ & \begin{tabular}[t]{@{}c@{}} -0.07 (0.06) \\ \relax [-0.0025] \end{tabular} & -- & -- & \begin{tabular}[t]{@{}c@{}} -0.01 (0.06) \\ \relax [-0.0003] \end{tabular} & \begin{tabular}[t]{@{}c@{}} -0.05 (0.06) \\ \relax [-0.0021] \end{tabular} \\
$\text{ROA}_{firm}$ & \begin{tabular}[t]{@{}c@{}} 0.00 (0.02) \\ \relax [0.0001] \end{tabular} & -- & -- & \begin{tabular}[t]{@{}c@{}} -0.01 (0.02) \\ \relax [-0.0003] \end{tabular} & \begin{tabular}[t]{@{}c@{}} -0.02 (0.01) \\ \relax [-0.0007] \end{tabular} \\
$\text{Tang}$ & \begin{tabular}[t]{@{}c@{}} 1.65*** (0.34) \\ \relax [0.0596] \end{tabular} & -- & -- & \begin{tabular}[t]{@{}c@{}} 0.14 (0.42) \\ \relax [0.0050] \end{tabular} & \begin{tabular}[t]{@{}c@{}} 0.31 (0.39) \\ \relax [0.0132] \end{tabular} \\
\midrule
$\ln t^{net}$ & -- & \begin{tabular}[t]{@{}c@{}} -0.17*** (0.04) \\ \relax [-0.0057] \end{tabular} & \begin{tabular}[t]{@{}c@{}} 0.44*** (0.02) \\ \relax [0.0179] \end{tabular} & \begin{tabular}[t]{@{}c@{}} -0.22*** (0.04) \\ \relax [-0.0075] \end{tabular} & \begin{tabular}[t]{@{}c@{}} 0.46*** (0.04) \\ \relax [0.0196] \end{tabular} \\
$\ln h$ & -- & \begin{tabular}[t]{@{}c@{}} 1.94*** (0.13) \\ \relax [0.0664] \end{tabular} & -- & \begin{tabular}[t]{@{}c@{}} 2.41*** (0.17) \\ \relax [0.0836] \end{tabular} & -- \\
$\ln t^{bal}$ & \begin{tabular}[t]{@{}c@{}} 0.58*** (0.10) \\ \relax [0.0210] \end{tabular} & -- & -- & \begin{tabular}[t]{@{}c@{}} -0.10 (0.10) \\ \relax [-0.0036] \end{tabular} & \begin{tabular}[t]{@{}c@{}} 0.53*** (0.08) \\ \relax [0.0228] \end{tabular} \\
$\ln\text{Assets}_{bank}$ & \begin{tabular}[t]{@{}c@{}} 0.38*** (0.09) \\ \relax [0.0138] \end{tabular} & -- & -- & \begin{tabular}[t]{@{}c@{}} -0.13 (0.10) \\ \relax [-0.0047] \end{tabular} & \begin{tabular}[t]{@{}c@{}} -0.59*** (0.07) \\ \relax [-0.0252] \end{tabular} \\
$\text{Lev}_{bank}$ & \begin{tabular}[t]{@{}c@{}} -0.06*** (0.02) \\ \relax [-0.0021] \end{tabular} & -- & -- & \begin{tabular}[t]{@{}c@{}} -0.00 (0.02) \\ \relax [-0.0000] \end{tabular} & \begin{tabular}[t]{@{}c@{}} 0.01 (0.01) \\ \relax [0.0003] \end{tabular} \\
$\text{ROA}_{bank}$ & \begin{tabular}[t]{@{}c@{}} -0.37** (0.17) \\ \relax [-0.0135] \end{tabular} & -- & -- & \begin{tabular}[t]{@{}c@{}} -0.01 (0.16) \\ \relax [-0.0005] \end{tabular} & \begin{tabular}[t]{@{}c@{}} -0.82*** (0.18) \\ \relax [-0.0351] \end{tabular} \\
\midrule
Observations & 9,379 & 9,379 & 9,379 & 9,379 & 9,379 \\
Pseudo $R^2$ & 0.328 & 0.394 & 0.283 & 0.380 & 0.249 \\
\hline
\hline
\multicolumn{6}{c}{\textbf{Panel B: Unconsolidated Sample}} \\
\hline
\hline
& {\bf Model 1} & \multicolumn{2}{c}{\bf Model 2} & \multicolumn{2}{c}{\bf Model 3} \\
\hline
Variable & (Gravity) & A (w/ degree) & B (w/o degree) & A (w/ degree) & B (w/o degree) \\
\hline
Intercept & -3.73*** (0.68) & -5.09*** (0.43) & -7.47*** (0.29) & -0.50 (1.05) & -1.42** (0.73) \\
\addlinespace
$\ln s^{net}$ & -- & \begin{tabular}[t]{@{}c@{}} 0.04 (0.03) \\ \relax [0.0013] \end{tabular} & \begin{tabular}[t]{@{}c@{}} -0.06*** (0.02) \\ \relax [-0.0024] \end{tabular} & \begin{tabular}[t]{@{}c@{}} -0.19*** (0.03) \\ \relax [-0.0065] \end{tabular} & \begin{tabular}[t]{@{}c@{}} -0.08*** (0.02) \\ \relax [-0.0034] \end{tabular} \\
$\ln k$ & -- & \begin{tabular}[t]{@{}c@{}} 1.56*** (0.11) \\ \relax [0.0480] \end{tabular} & -- & \begin{tabular}[t]{@{}c@{}} \textbf{1.76*** (0.12)} \\ \relax \ [0.0599] \end{tabular} & -- \\
$is\_exclusive$ & -- & \begin{tabular}[t]{@{}c@{}} 6.59*** (0.61) \\ \relax [0.2023] \end{tabular} & -- & \begin{tabular}[t]{@{}c@{}} 0.86* (0.44) \\ \relax [0.0292] \end{tabular} & -- \\
$\ln s^{bal}$ & \begin{tabular}[t]{@{}c@{}} -0.06*** (0.02) \\ \relax [-0.0025] \end{tabular} & -- & -- & \begin{tabular}[t]{@{}c@{}} -0.13*** (0.02) \\ \relax [-0.0045] \end{tabular} & \begin{tabular}[t]{@{}c@{}} -0.08*** (0.02) \\ \relax [-0.0034] \end{tabular} \\
$\ln\text{Assets}_{firm}$ & \begin{tabular}[t]{@{}c@{}} -0.67*** (0.05) \\ \relax [-0.0298] \end{tabular} & -- & -- & \begin{tabular}[t]{@{}c@{}} 0.00 (0.05) \\ \relax [0.0000] \end{tabular} & \begin{tabular}[t]{@{}c@{}} -0.40*** (0.04) \\ \relax [-0.0174] \end{tabular} \\
$\text{Lev}_{firm}$ & \begin{tabular}[t]{@{}c@{}} 0.00 (0.00) \\ \relax [0.0001] \end{tabular} & -- & -- & \begin{tabular}[t]{@{}c@{}} 0.00 (0.00) \\ \relax [0.0001] \end{tabular} & \begin{tabular}[t]{@{}c@{}} 0.00*** (0.00) \\ \relax [0.0002] \end{tabular} \\
$\text{ROA}_{firm}$ & \begin{tabular}[t]{@{}c@{}} 0.02 (0.01) \\ \relax [0.0007] \end{tabular} & -- & -- & \begin{tabular}[t]{@{}c@{}} 0.01 (0.01) \\ \relax [0.0003] \end{tabular} & \begin{tabular}[t]{@{}c@{}} 0.01 (0.01) \\ \relax [0.0004] \end{tabular} \\
$\text{Tang}$ & \begin{tabular}[t]{@{}c@{}} -1.07*** (0.16) \\ \relax [-0.0473] \end{tabular} & -- & -- & \begin{tabular}[t]{@{}c@{}} -0.19 (0.18) \\ \relax [-0.0064] \end{tabular} & \begin{tabular}[t]{@{}c@{}} -0.72*** (0.16) \\ \relax [-0.0311] \end{tabular} \\
\midrule
$\ln t^{net}$ & -- & \begin{tabular}[t]{@{}c@{}} -0.64*** (0.05) \\ \relax [-0.0197] \end{tabular} & \begin{tabular}[t]{@{}c@{}} 0.44*** (0.02) \\ \relax [0.0186] \end{tabular} & \begin{tabular}[t]{@{}c@{}} -0.66*** (0.05) \\ \relax [-0.0224] \end{tabular} & \begin{tabular}[t]{@{}c@{}} 0.27*** (0.03) \\ \relax [0.0116] \end{tabular} \\
$\ln h$ & -- & \begin{tabular}[t]{@{}c@{}} 2.88*** (0.13) \\ \relax [0.0885] \end{tabular} & -- & \begin{tabular}[t]{@{}c@{}} 2.90*** (0.15) \\ \relax [0.0987] \end{tabular} & -- \\
$\ln t^{bal}$ & \begin{tabular}[t]{@{}c@{}} 0.12 (0.08) \\ \relax [0.0055] \end{tabular} & -- & -- & \begin{tabular}[t]{@{}c@{}} -0.45*** (0.09) \\ \relax [-0.0155] \end{tabular} & \begin{tabular}[t]{@{}c@{}} 0.32*** (0.07) \\ \relax [0.0137] \end{tabular} \\
$\ln\text{Assets}_{bank}$ & \begin{tabular}[t]{@{}c@{}} 0.53*** (0.07) \\ \relax [0.0236] \end{tabular} & -- & -- & \begin{tabular}[t]{@{}c@{}} 0.39*** (0.09) \\ \relax [0.0134] \end{tabular} & \begin{tabular}[t]{@{}c@{}} -0.11 (0.08) \\ \relax [-0.0047] \end{tabular} \\
$\text{Lev}_{bank}$ & \begin{tabular}[t]{@{}c@{}} -0.03*** (0.01) \\ \relax [-0.0013] \end{tabular} & -- & -- & \begin{tabular}[t]{@{}c@{}} 0.00 (0.01) \\ \relax [0.0001] \end{tabular} & \begin{tabular}[t]{@{}c@{}} -0.02* (0.01) \\ \relax [-0.0009] \end{tabular} \\
$\text{ROA}_{bank}$ & \begin{tabular}[t]{@{}c@{}} -0.10 (0.11) \\ \relax [-0.0043] \end{tabular} & -- & -- & \begin{tabular}[t]{@{}c@{}} -0.25 (0.18) \\ \relax [-0.0086] \end{tabular} & \begin{tabular}[t]{@{}c@{}} -0.57*** (0.12) \\ \relax [-0.0248] \end{tabular} \\
\midrule
Observations & 11,200 & 11,200 & 11,200 & 11,200 & 11,200 \\
Pseudo $R^2$ & 0.201 & 0.415 & 0.216 & 0.396 & 0.198 \\
\hline
\hline
\end{tabular*}}
\caption{Step $\#1$: Determinants of link formation: Model 1 tests fundamental variables, Model 2 tests network variables, Model 3 tests both. Models 2B and 3B exclude the degree to test for omitted variable bias. Coefficients are reported with standard errors in parentheses. Significance: *** $p<0.01$, ** $p<0.05$, * $p<0.1$. Values in square brackets $[\dots]$ represent Average Marginal Effects (AME).}
\label{tab:logit_results}
\end{table*}

\section{Results}

In the first stage, we employ a logistic regression to investigate the determinants of link formation. The results, reported in Table~\ref{tab:logit_results}, reveal that the degree remains the dominant predictor of new relationships in both samples: this suggests the presence of preferential attachment, i.e. banks select clients based on their existing connectivity, effectively interpreting a firm's number of lenders as a certification of quality. Crucially, the sign reversal of size variables when degree is included (Model 3) highlights that the topological `herd' effect, rather than the absolute financial volume, drives access to new credit. Furthermore, the inclusion of the \emph{is\_exclusive} dummy shows that while multi-banked firms enjoy easier access to new lenders, single-banked firms (exclusive) face a significantly different entry barrier, particularly in the unconsolidated sample.

In the second stage, we investigate the determinants of loan size. Results, reported in Table~\ref{tab:ols_results}, document a dual mechanism: while balance sheet capacity ($s^{bal}$) and network strength ($s^{net}$) determine the overall credit `ceiling', firms' degree has a robust negative effect. In the consolidated sample, we find a degree elasticity of, approximately, $-1.04$: this implies that while firms strategically diversify to mitigate the `hold-up' risk, there is a structural friction, i.e. adding relationships does not lead to a proportional increase in total credit. The individual loan size per bank, instead, drops sharply, suggesting that coordination costs and administrative overhead may lead to credit stagnation as the network structure becomes overly complex.

A key finding regarding the hierarchy of credit determinants emerges from the role of asset tangibility. In the fundamental-only specification (Model 1), tangibility is a strong predictor of loan size for large corporate groups. However, as shown in Table~\ref{tab:ols_results}, once network variables are introduced (Model 3A), the significance of tangibility vanishes across both samples: this suggests a total `network substitution effect', i.e. the reputational collateral provided by a firm's position in the credit network - particularly, its observed network strength - effectively overrides the role of physical collateral. Even for large consolidated groups, the collective validation of the banking system appears more informative for loan sizing than the liquidation value of tangible assets.

Overall, our multi-stage regression analysis reveals that network topology and financial fundamentals play divergent roles for credit allocation: while network topology is the primary determinant of \emph{credit access} (link formation), the \emph{credit volume} (loan sizing) is driven by a sophisticated dual-signaling mechanism.\\

We also document a striking divergence between reported balance sheet debt ($s^{bal}$) and observed network strength ($s^{net}$) for the unconsolidated sample: as shown in Table~\ref{tab:ols_results} (Model 3A), $s^{bal}$ exerts a significantly negative effect on loan size ($\beta \approx -0.09^{***}$), while $s^{net}$ remains a strong positive predictor ($\beta \approx 0.53^{***}$). This result provides a deep reflection of the information asymmetry inherent in the SME market: for opaque firms, high debt on the balance sheet is interpreted by lenders as a traditional risk signal (over-indebtedness or distress). Conversely, the `network footprint' ($s^{net}$) acts as a powerful countervailing signal of market validation; banks are willing to grant larger loans only when they see that the firm's debt is actually `backed' by the collective trust of other reporting lenders. In contrast, for large consolidated groups, $s^{bal}$ becomes non-significant, as these entities possess higher transparency and alternative ways to signal solvency.

Furthermore, we document a consistently negative effect of the degree on individual loan size. For the consolidated sample, the magnitude of degree elasticity ($-1.04$) suggests a potential contraction in total credit volume as relationships multiply, but we interpret this result with scientific caution given the observed standard error ($0.15$). Specifically, while the negative sign is highly robust and confirms a strategic diversification strategy where firms fragment debt to mitigate hold-up risks, the proximity to unit elasticity indicates that large groups are operating at a threshold of structural friction. At this stage, adding new banking relationships leads to a sharp drop in individual loan sizes that nearly offsets the marginal gain in access. This patterns provides a strong signal of a \emph{complexity penalty} consistent with coordination failure theories, although longitudinal data would be required to definitively establish whether over-diversification leads to an absolute stagnation in total credit depth.

To ensure the stability of these estimates, we performed a multicollinearity check (reported in Appendix~\hyperlink{AppB}{B}), which reveals high Variance Inflation Factors ($\text{VIF}>10$) for bank-side variables, reported in Table~\ref{tab:vif}. To address this, we re-estimated the loan sizing model using Bank Fixed Effects to absorb all time-invariant bank heterogeneity. The results in Table~\ref{tab:fixed_effects} corroborate the main analysis: even within the same bank, network exposure remains the dominant driver of loan volume, while the diversification constraint provides a significantly negative friction.

\begin{table*}[t!]
\centering
\resizebox{\textwidth}{!}{%
\begin{tabular*}{\textwidth}{@{\extracolsep{\fill}}lccccc}
\hline
\hline
\multicolumn{6}{c}{\bf Panel A: Consolidated Sample} \\
\hline
\hline
& {\bf Model 1} & \multicolumn{2}{c}{\bf Model 2} & \multicolumn{2}{c}{\bf Model 3} \\
\hline
Variable & (w/o degree) & A (w/ degree) & B (w/o degree) & A (w/ degree) & B (w/o degree) \\
\hline
Intercept & -3.80*** (1.20) & 0.39 (0.58) & 4.51*** (0.62) & -5.35*** (1.62) & -2.76* (1.48) \\
\addlinespace
$\ln s^{net}$ & -- & 0.77*** (0.05) & 0.16*** (0.05) & 0.71*** (0.06) & 0.13*** (0.05) \\
$\ln k$ & -- & -1.18*** (0.14) & -- & \textbf{-1.04*** (0.15)} & -- \\
$is\_exclusive$ & -- & 7.40*** (0.69) & -- & \textbf{6.93*** (0.75)} & -- \\
$\ln s^{bal}$ & 0.13** (0.05) & -- & -- & 0.02 (0.05) & 0.05 (0.06) \\
$\ln\text{Assets}_{firm}$ & 0.15** (0.06) & -- & -- & 0.04 (0.04) & 0.16*** (0.06) \\
$\text{Lev}_{firm}$ & 0.12 (0.10) & -- & -- & -0.01 (0.09) & 0.13 (0.10) \\
$\text{ROA}_{firm}$ & -0.03** (0.02) & -- & -- & -0.00 (0.02) & -0.03** (0.01) \\
$\text{Tang}$ & 1.17*** (0.43) & -- & -- & 0.43 (0.42) & 0.95** (0.43) \\
\midrule
$\ln t^{net}$ & -- & 0.09* (0.05) & 0.24*** (0.02) & 0.07 (0.05) & 0.09** (0.04) \\
$\ln h$ & -- & 0.34** (0.16) & -- & -0.09 (0.21) & -- \\
$\ln t^{bal}$ & -0.10 (0.13) & -- & -- & 0.02 (0.11) & -0.16 (0.13) \\
$\ln\text{Assets}_{bank}$ & 0.61*** (0.13) & -- & -- & 0.35*** (0.12) & 0.50*** (0.14) \\
$\text{Lev}_{bank}$ & 0.02 (0.03) & -- & -- & 0.03 (0.02) & 0.03 (0.03) \\
$\text{ROA}_{bank}$ & 0.04 (0.32) & -- & -- & -0.14 (0.26) & 0.00 (0.30) \\
\midrule
Observations & 486 & 486 & 486 & 486 & 486 \\
$R^2$ & 0.294 & 0.468 & 0.226 & 0.491 & 0.328 \\
\hline
\hline
\multicolumn{6}{c}{\bf Panel B: Unconsolidated Sample} \\
\hline
\hline
& {\bf Model 1} & \multicolumn{2}{c}{\bf Model 2} & \multicolumn{2}{c}{\bf Model 3} \\
\hline
Variable & (w/o degree) & A (w/ degree) & B (w/o degree) & A (w/ degree) & B (w/o degree) \\
\hline
Intercept & -4.81*** (1.13) & 2.47*** (0.52) & 6.74*** (0.45) & -4.44*** (1.49) & -5.55*** (1.36) \\
\addlinespace
$\ln s^{net}$ & -- & 0.67*** (0.05) & 0.10*** (0.03) & 0.53*** (0.06) & 0.12*** (0.03) \\
$\ln k$ & -- & -0.97*** (0.14) & -- & \textbf{-0.53*** (0.15)} & -- \\
$is\_exclusive$ & -- & 6.42*** (0.60) & -- & \textbf{5.00*** (0.67)} & -- \\
$\ln s^{bal}$ & -0.06*** (0.02) & -- & -- & -0.09*** (0.02) & -0.11*** (0.02) \\
$\ln\text{Assets}_{firm}$ & 0.64*** (0.07) & -- & -- & 0.34*** (0.08) & 0.60*** (0.07) \\
$\text{Lev}_{firm}$ & 0.00* (0.00) & -- & -- & 0.00 (0.00) & 0.00 (0.00) \\
$\text{ROA}_{firm}$ & 0.01 (0.01) & -- & -- & 0.01 (0.01) & 0.01 (0.01) \\
$\text{Tang}$ & 0.34 (0.24) & -- & -- & 0.33 (0.23) & 0.33 (0.25) \\
\midrule
$\ln t^{net}$ & -- & 0.01 (0.04) & 0.13*** (0.02) & -0.02 (0.04) & -0.00 (0.03) \\
$\ln h$ & -- & 0.30** (0.15) & -- & 0.12 (0.18) & -- \\
$\ln t^{bal}$ & -0.28*** (0.06) & -- & -- & -0.27*** (0.05) & -0.28*** (0.06) \\
$\ln\text{Assets}_{bank}$ & 0.58*** (0.08) & -- & -- & 0.50*** (0.11) & 0.61*** (0.10) \\
$\text{Lev}_{bank}$ & 0.02 (0.02) & -- & -- & 0.03* (0.02) & 0.03 (0.02) \\
$\text{ROA}_{bank}$ & -0.02 (0.22) & -- & -- & -0.17 (0.20) & -0.10 (0.22) \\
\midrule
Observations & 575 & 575 & 575 & 575 & 575 \\
$R^2$ & 0.302 & 0.361 & 0.097 & 0.442 & 0.337 \\
\hline
\hline
\end{tabular*}}
\caption{Step $\#2$: Determinants of loan size: Model 1 tests fundamental variables, Model 2 tests network variables, Model 3 tests both. Models 2B and 3B exclude degree centrality to test for omitted variable bias. Coefficients are reported with standard errors in parentheses. Significance: *** $p<0.01$, ** $p<0.05$, * $p<0.1$.}
\label{tab:ols_results}
\end{table*}

\subsection{Robustness check: the null model placebo}

To prove that the observed topological effects represent strategic behavioural choices rather than random numerical consequences of node size, we performed a `placebo test' by replacing empirical degrees with those predicted by our maximum-entropy benchmark (\emph{Null Net} and \emph{Null Bal}).

Methodologically, we avoid circularity through a `cross-controlling' procedure: the \emph{Null Net} model (which uses observed $s^{net}$ as fitness) is controlled for $s^{bal}$ in the regressions, while the \emph{Null Bal} model (derived from $s^{bal}$) is controlled for $s^{net}$. Specifically, this allows us to distinguish between two different signaling mechanisms: while the \emph{Null Net} model tests whether connectivity is a mere byproduct of a firm's total credit volume (volume-driven fitness), the \emph{Null Bal} model identifies if it stems from the firm's accounting size (balance-sheet-driven fitness). By isolating these effects, we can determine if empirical connectivity carries a distinct reputational signal that exceeds what either form of `mass' would statistically imply.

As shown in Table~\ref{tab:logit_robustness}, the empirical results for credit access vastly outperform the null benchmarks. In the link-formation stage, the null models tend to predict a `saturation effect': for large firms, the probability of forming new links often decreases or stagnates in the counterfactual world because their capacity is already statistically exhausted by their size. In stark contrast, the real Italian market treats existing connectivity as a strong positive signal. The fact that the empirical preferential attachment is significantly more intense than the null expectation confirms that banks utilize the topology as a certification of creditworthiness that overrides the statistical constraints of firm size.

The most compelling evidence, however, emerges from the loan sizing analysis (Table~\ref{tab:ecapm_regression}). We observe a striking sign reversal for firms between the counterfactual and empirical worlds. In both benchmarks, the coefficient associated with the degree of firms is strongly positive. This represents the expected `size effect': in a random world, bigger firms naturally have both more bank connections and larger loans; conversely, in the empirical network, the coefficient associated with the degree of firms is robustly negative. This divergence provides definitive proof that debt fragmentation is a borrower-side strategic choice in the Italian market. While the benchmark reflects a world where connectivity is a by-product of fitnesses, the real-world negative correlation isolates a deliberate strategy where firms fragment their debt to balance bargaining power against coordination frictions. Regarding the banking side, while we control for supply-side heterogeneity, the high collinearity between bank-level size metrics prevents a clear identification of specific lending strategies; our primary and most robust evidence remains centered on the borrower’s structural response to informational opacity.

\newpage

\section{The Case for Topological Supervision:\\Network Dynamics and Financial Stability}

The empirical evidence presented in this study suggests that the Italian corporate lending market operates as a complex information-processing system where topological signals are often more relevant than individual financial indicators: this shift from `hard information' (balance sheets) to `network information' (connectivity) introduces a specific kind of systemic risk that fall directly within the mandate of macro-prudential supervision. We argue that proactive oversight is required to address the structural vulnerabilities emerging from the interplay between bank-herding and firm-diversification strategies.

A primary concern for financial stability is the documented preferential attachment effect in the link-formation stage. This suggests a systemic reliance on `topological certification', where lenders interpret the existing connectivity of a borrower as a proxy for creditworthiness: while leveraging the collective monitoring of the system can reduce individual screening costs~\cite{Carletti:2007aa}, it risks creating informational bubbles. When secondary and tertiary lenders `follow the herd', the market may enter a screening vacuum characterized by a free-riding problem. As noted in the literature on financial contagion~\cite{Allen:2000aa}, if a lead bank's initial due diligence is flawed, this error can propagate through the entire network cluster without being caught by subsequent participants who assume the borrower has already been well-validated. Supervisory authorities should therefore move beyond bilateral solvency checks and begin monitoring `topological herding' to ensure that banks maintain independent screening standards even when entering well-validated clusters.

Moreover, the strategic diversification identified in our OLS results - while rational for individual firms seeking to mitigate `hold-up' risk~\cite{Sharpe:1990aa,Rajan:1992aa} - introduces significant coordination failure risk at the system level. In the event of a borrower’s financial distress, a hyper-fragmented debt structure becomes a formidable barrier to efficient restructuring. The transaction costs and legal complexities of coordinating a large pool of heterogeneous lenders often lead to `liquidity pull-backs' and unnecessary insolvencies \cite{Bolton:1996aa}. For a supervisor, this necessitates a shift in focus from traditional `concentration risk' (the risk of having too much exposure to one client) to `granularity risk' (the risk of having too many lenders for a single client). Financial stability is compromised when a debt structure becomes so fragmented that no single `lead bank' maintains enough `skin-in-the-game' to act as an effective coordinator during a recovery phase~\cite{Mistrulli:2011aa}.

Finally, the documented degree elasticity of $-1.04$ for large corporate groups points to a fundamental systemic inefficiency. This `complexity penalty' suggests that as banking relationships multiply, the marginal utility of new connections vanishes, leading to a stagnation in credit depth. From a macro-prudential perspective, this indicates that the administrative and coordination overhead of maintaining an overly complex network acts as a structural friction that hampers capital allocation. Furthermore, this fragmentation likely inflates the system-wide Loss Given Default (LGD). As noted by Davydenko and Franks (2008), when a borrower defaults on a hyper-fragmented debt structure, the recovery process is plagued by high coordination costs and legal frictions among a multitude of small creditors, which significantly erodes the reorganization value of the firm~\cite{Davydenko:2008aa}. As suggested by Haldane (2009) in the context of network complexity and stability~\cite{Haldane:2009aa}, such a configuration increases the fragility of the system: the debt is spread so thinly that the banking system loses its ability to provide meaningful support to firms during downturns. Policy interventions should therefore aim to foster an efficient middle ground, ensuring that the credit market provides sufficient funding depth without forcing firms into hyper-fragmented topological structures that ultimately increase systemic fragility and reduce the effectiveness of relationship lending.

Beyond structural frictions, the digital transformation of financial intermediation introduces new layers of systemic vulnerability. As banks increasingly move toward AI-driven credit scoring, there is a looming risk of \textit{Algorithmic Herding}: if machine-learning models are trained on topological signals - essentially automated versions of the preferential attachment we document - the credit market risks becoming a self-referential echo chamber. Our results provide a blueprint for a new generation of \emph{topological stress tests}, allowing supervisors to simulate how shocks propagate not through balance-sheet contagion but through the collapse of topological trust clusters.

\section{Discussion}

Our investigation of the Italian bank-firm credit market reveals that the allocation of credit is driven less by the intrinsic financial health of borrowers and more by the topological structure of the lending network itself. By constructing a novel dataset of bilateral exposures and benchmarking it against an entropy-based null model, we have isolated the specific role of network signals in credit decisions.

Our findings support three main conclusions regarding the mechanics of credit allocation. First, the credit market operates as a reputational engine where degree centrality acts as a gatekeeper. Link formation follows a preferential attachment dynamic that is strictly topological; once a firm establishes connections, it attracts more lenders regardless of its profitability or leverage. This suggests that banks leverage the screening efforts of incumbent lenders, interpreting existing relationships as a `topological certification' of quality, facilitated by information-sharing systems like the \emph{Centrale Rischi}~\cite{Pagano:1993aa}.

Second, we identify a profound `network substitution effect' that redefines the role of collateral. While traditional theory emphasizes the importance of physical assets, our full competition model shows that the significance of asset tangibility vanishes across both corporate segments when network variables are included. This implies that the reputational collateral provided by a firm's network footprint, i.e. $s^{net}$, effectively overrides the liquidation value of fixed assets. This substitution is particularly evident for opaque SMEs, where reported balance sheet debt, i.e. $s^{bal}$, is interpreted as a risk signal (negative coefficient), while observed network strength acts as the primary form of market validation.

Third, our counterfactual analysis proves that this structure is a \emph{borrower-side strategic choice} rather than a random artifact of firm size. The striking sign reversal in the OLS regressions - where empirical degree negatively impacts loan size for firms while the null model predicts a strong positive correlation - highlights that Italian firms actively manage their network positions to mitigate `hold-up' risks. Specifically, our cross-controlling procedure allows us to distinguish between two signaling mechanisms: while the \emph{Null Net} model tests whether connectivity is a mere by-product of a firm's total credit volume (volume-driven fitness), the \emph{Null Bal} model identifies if it stems from the firm's accounting size (balance-sheet-driven fitness). By isolating these effects, we confirm that empirical connectivity carries a distinct reputational signal that exceeds what either form of `mass' would statistically imply. This strategic fragmentation, while defensive for the borrower, creates structural vulnerabilities: the transaction costs and legal complexities of coordinating a large pool of lenders often lead to coordination failure during financial distress~\cite{Bolton:1996aa}. It should be noted that while we control for supply-side factors, the high collinearity between bank-level size metrics prevents a definitive identification of lending strategies; thus, our most robust evidence remains centered on the firms' strategic response to informational opacity.

This finding challenges the long-standing regulatory dogma that credit diversification is an unambiguous good. While traditional portfolio theory suggests that spreading risk across multiple lenders enhances stability, we reveal a \emph{diversification paradox}: beyond a certain threshold of complexity, the marginal gain in credit access is offset by coordination frictions and monitoring free-riding. This suggests that the Italian market may be trapped in a `hyper-fragmentation equilibrium' where systemic stability is traded off for individual bargaining power, ultimately leading to a more fragile financial architecture in times of crisis.

The degree elasticity for large groups is robustly negative and near the unit threshold, providing a strong signal of a potential `complexity penalty'. Although the observed standard error suggests caution in establishing an absolute contraction in total credit volume, the results indicate that over-diversification leads to a stagnation in credit depth and a likely increase in the system-wide Loss Given Default (LGD)~\cite{Davydenko:2008aa}. These results call for a shift toward `topological supervision'. As credit allocation increasingly relies on network signals rather than hard information, macro-prudential authorities must monitor `topological herding' and `granularity risk' to identify structural vulnerabilities - such as restructuring bottlenecks or informational bubbles~\cite{Allen:2000aa, Haldane:2009aa} - that traditional bilateral indicators are fundamentally unable to capture.

Nonetheless, we acknowledge that the evidence presented here is based on a cross-sectional snapshot of the 2022 fiscal year. While this snapshot offers compelling insights into the current state of the Italian market, more empirical evidence is required to put our findings on a more firm ground. Future research should aim for a more rigorous longitudinal analysis, extending the scope to cover multiple years and incorporating even smaller firms. Such a broader investigation would be desirable to confirm the persistence of the `topological certification' and `complexity penalty' effects across different macroeconomic cycles and further validate the role of network structure as a primary determinant of financial intermediation.

Ultimately, our research signals the \emph{twilight of the balance sheet} as the primary anchor of corporate lending. In the modern financial ecosystem, the network is not just a map of transactions: it is the fundamental infrastructure of creditworthiness; ignoring these topological currents is no longer an option for supervisors tasked with maintaining the stability of a complex, interconnected global economy.

\section{Acknowledgments}

This work has been supported by the following projects: `C2T - From Crises to Theory: towards a science of resilience and recovery for economic and financial systems' - P2022E93B8, Funded by the European Union Next Generation EU, PNRR Mission 4 Component 2 Investment 1.1, CUP: D53D23019330001; `RE-Net - Reconstructing economic networks: from physics to machine learning and back' - 2022MTBB22, Funded by the European Union Next Generation EU, PNRR Mission 4 Component 2 Investment 1.1, CUP: D53D23002330006.

\section{Author contributions}

Study conception and design: TS, GC. Data collection: AP, AM. Analysis and interpretation of results: AP, AM, TS, GC. Draft manuscript preparation and revision: AP, AM, TS, GC. 

\section{Competing interests}

The authors declare no competing interests.

\appendix

\section*{APPENDIX A.\\Derivation of the maximum-entropy counterfactual}
\hypertarget{AppA}{}

We want to build a probability distribution $P(\mathbf{G})$ on the ensemble $\Omega$ of all possible bipartite networks $\mathbf{G}$, defined by $N_F$ firms and $N_B$ banks, by fixing the expected value of some network properties (the so-called \emph{constraints}) while randomizing everything else. As prescribed by Information Theory~\cite{PhysRev.106.620}, this is obtained by maximizing the Shannon entropy $S=-\sum_{\mathbf{G}\in\Omega}P(\mathbf{G})\ln P(\mathbf{G})$ in a constrained fashion~\cite{Park:2004aa,Cimini:2019aa}. 

The \textit{Bipartite Configuration Model} (BiCM)~\cite{Saracco:2015aa} is obtained by constraining the degree of each node: upon doing so, $P(\mathbf{G})$ factorizes into link-specific terms, the one representing the connection probability between firm $i$ and bank $j$ reading

\begin{equation}
p_{ij}=\frac{x_i y_j}{1 + x_i y_j};
\end{equation}
the Lagrange multipliers $x_i$ and $y_j$ are determined by solving the system of coupled equations $\sum_{j=1}^{N_B}p_{ij}=k_i$ and $\sum_{i=1}^{N_F}p_{ij}=h_j$~\cite{Saracco:2015aa}.

Whenever the information about degrees is not available, \emph{fitness-based models}~\cite{Caldarelli:2002aa} can be used as heuristic alternatives: they assume the same functional form of the link probability but replace the Lagrange multipliers with known quantities representing the `intrinsic capacity' of nodes to form connections: to this aim, node strengths are typically used, thus leading to the formula defining our benchmark~\cite{Cimini:2015aa,Squartini:2017aa}

\begin{equation}
p_{ij}=\frac{zs_it_j}{1+zs_it_j};
\end{equation}
the global parameter $z$ is calibrated to ensure that the expected number of links, say $\langle L\rangle$, matches the empirical one, say $L_{obs}$: in formulas,

\begin{equation}
\langle L\rangle=\sum_{i=1}^{N_F}\sum_{j=1}^{N_B}\frac{zs_it_j}{1+zs_it_j}=L_{obs}.
\end{equation}

Concerning link weights, a rigorous maximum-entropy prescription would consists in constraining both degrees and strengths - a recipe that has been named \emph{Enhanced Configuration Model} (ECM)~\cite{Mastrandrea:2014aa,Gabrielli:2019aa}. A heuristic, but efficient, alternative consists in considering the formulation named \emph{density-corrected Gravity Model} (dcGM)~\cite{Cimini:2015aa,Squartini:2017aa}, the one we employ here:

\begin{equation}
w_{ij|_{a_{ij}=1}}=\frac{s_it_j}{Wp_{ij}};
\label{gravity}
\end{equation}
naturally, $W=\sum_{i=1}^{N_F}s_i=\sum_{j=1}^{N_B}t_j$ whenever the system is closed. We redirect the interested reader to the publications~\cite{Cimini:2015aa,Squartini:2017aa} for further details.

\newpage

\section*{APPENDIX B.\\Diagnostic Tests}
\hypertarget{AppB}{}

\begin{table}[h!]
\centering
\begin{tabular*}{\textwidth}{@{\extracolsep{\fill}}lcc}
\hline
\hline
{\bf Variable} & {\bf Consolidated} & {\bf Unconsolidated} \\
\hline
\hline
$\ln h$ & 12.82 & 11.08 \\
$\ln \text{Assets}_{bank}$ & 11.06 & 9.85 \\
$\ln t^{bal}$ & 10.77 & 4.63 \\
$\ln t^{net}$ & 7.81 & 6.99 \\
$\ln s^{net}$ & 5.12 & 7.05 \\
$is\_exclusive$ & 3.53 & 4.85 \\
$\ln k$ & 2.17 & 2.71 \\
$\ln \text{Assets}_{firm}$ & 1.78 & 1.55 \\
$\text{Lev.}_{firm}$ & 1.49 & 1.25 \\
$\ln s^{bal}$ & 1.46 & 1.40 \\
$\text{Tang.}$ & 1.22 & 1.09 \\
$\text{ROA}_{firm}$ & 1.19 & 1.27 \\
$\text{Lev.}_{bank}$ & 1.11 & 1.14 \\
$\text{ROA}_{bank}$ & 1.15 & 1.10 \\
\hline
\hline
\end{tabular*}
\caption{Variance Inflation Factors (VIF) for Model 3 variables calculated after the `full Herman correction'. High VIF values for bank-side variables ($\ln h_{bank}$, $\ln \text{Assets}_{bank}$, $\ln t^{bal}_{bank}$) justify the use of bank fixed effects to absorb supply-side collinearity.}
\label{tab:vif}
\end{table}

\begin{table}[h!]
\centering
\begin{tabular*}{\textwidth}{@{\extracolsep{\fill}}lcc}
\hline
\hline
{\bf Variable} & {\bf Consolidated} & {\bf Unconsolidated} \\
\hline
\hline
$\ln s^{net}$ & 0.70*** (0.07) & 0.56*** (0.07) \\
$\ln k$ & -0.82*** (0.16) & -0.42** (0.17) \\
$is\_exclusive$ & 6.81*** (0.78) & 5.10*** (0.71) \\
$\ln s^{bal}$ & 0.03 (0.05) & -0.09*** (0.02) \\
$\ln \text{Assets}_{firm}$ & 0.03 (0.05) & 0.29*** (0.08) \\
$\text{Lev.}_{firm}$ & -0.01 (0.09) & 0.00 (0.00) \\
$\text{ROA}_{firm}$ & -0.00 (0.01) & 0.01 (0.02) \\
$\text{Tang.}$ & 0.48 (0.40) & 0.44* (0.26) \\
\midrule
Bank Controls & \textit{Absorbed} & \textit{Absorbed} \\
Bank Fixed Effects & \textit{Yes} & \textit{Yes} \\ \hline
Observations & 492 & 582 \\
$R^2$ & 0.594 & 0.512 \\
\hline \hline
\end{tabular*}
\caption{Loan sizing regression with bank fixed effects after the `full Herman correction'. This specification absorbs all bank-level heterogeneity and potential multicollinearity identified in the VIF analysis. Coefficients are reported with standard errors in parentheses. Significance: *** $p<0.01$, ** $p<0.05$, * $p<0.1$.}
\label{tab:fixed_effects}
\end{table}

\begin{figure*}[t!]
\centering
\includegraphics[width = \textwidth]{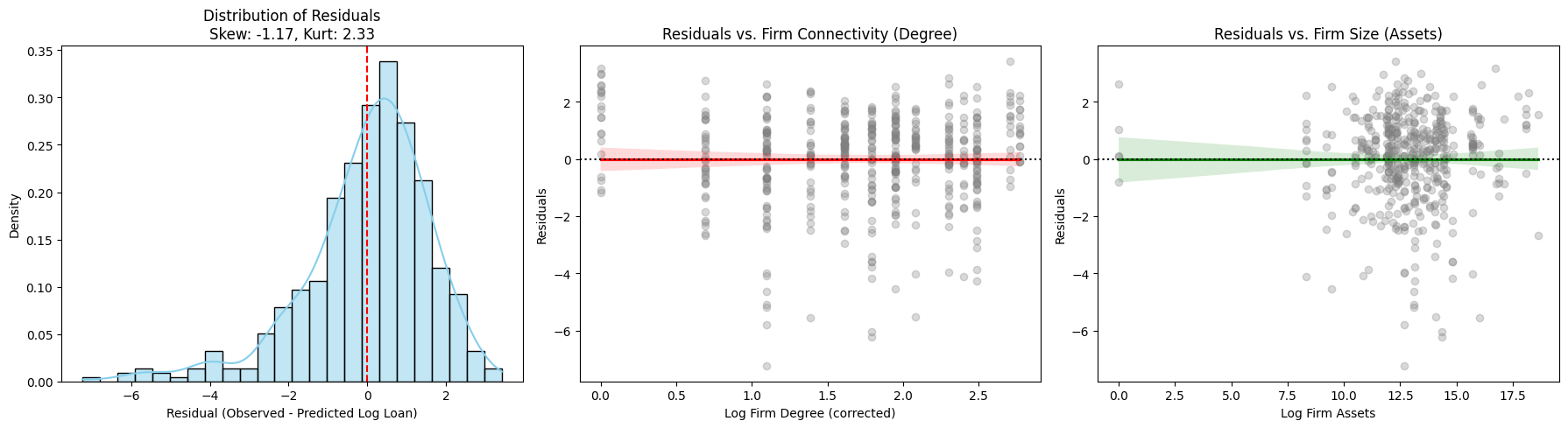}
\includegraphics[width = \textwidth]{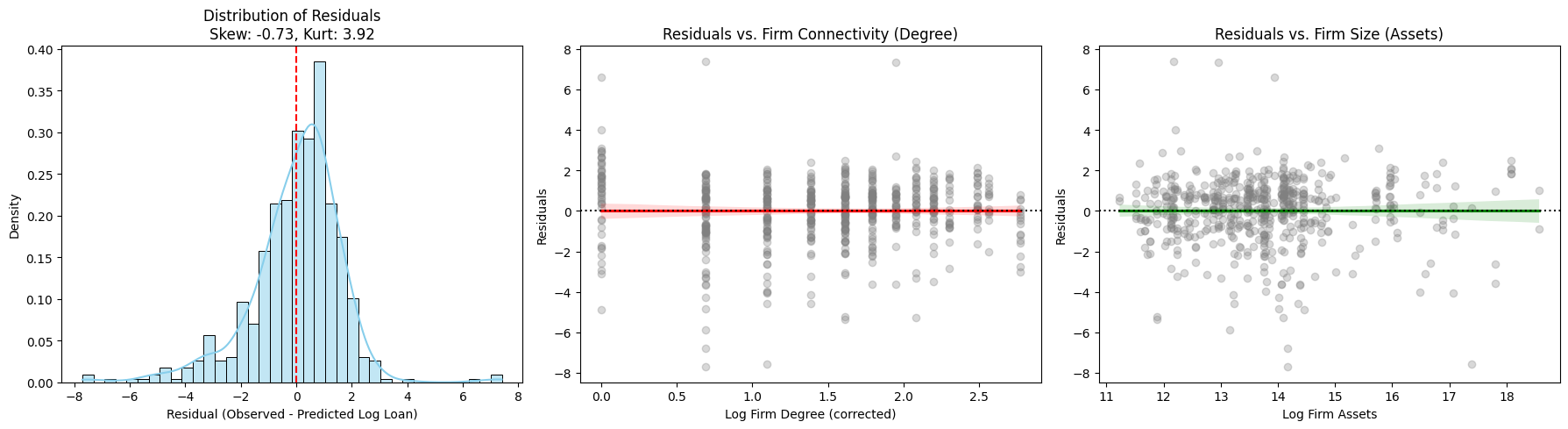}
\caption{Top panels: diagnostic analysis of residuals from the OLS Model 3 for the consolidated sample. The distribution (left) highlights negative skewness, while scatter plots visualize the relationship between residuals and firm connectivity (center) and size (right). Bottom panels: diagnostic analysis of residuals from the OLS Model 3 for the unconsolidated sample. Note the higher kurtosis (`fat tails') compared to the consolidated sample.}
\label{fig:residuals_cons}
\end{figure*}

\clearpage

\begin{table*}[!ht]
\centering
\renewcommand{\arraystretch}{0.9} % Riduce lo spazio tra le righe per far rientrare la tabella
\setlength{\tabcolsep}{3pt}       % Riduce lo spazio tra le colonne
\footnotesize
\begin{tabular*}{\textwidth}{@{\extracolsep{\fill}}lcccc}
\hline
\hline
\multicolumn{5}{c}{\bf Panel A: Consolidated Sample}\\
\hline
\hline
Variable & 1. Empirical & 2. Emp. (No $s^{net}$, $t^{net}$) & 3. Null Net & 4. Null Bal \\ \hline
Intercept & \begin{tabular}[t]{@{}c@{}} -0.6276 \\ (1.0698) \end{tabular} & \begin{tabular}[t]{@{}c@{}} -0.4530 \\ (1.0291) \end{tabular} & \begin{tabular}[t]{@{}c@{}} -0.3633 \\ (1.0724) \end{tabular} & \begin{tabular}[t]{@{}c@{}} -1.0713 \\ (1.1455) \end{tabular} \\
$\ln s^{net}$ & \begin{tabular}[t]{@{}c@{}} -0.1653*** (0.0391) \\ \relax [-0.0057] \end{tabular} & -- & -- & \begin{tabular}[t]{@{}c@{}} 0.0209 (0.0286) \\ \relax [0.0008] \end{tabular} \\
$\ln k$ & \begin{tabular}[t]{@{}c@{}} 1.5522*** (0.1210) \\ \relax [0.0539] \end{tabular} & \begin{tabular}[t]{@{}c@{}} 1.2541*** (0.1034) \\ \relax [0.0458] \end{tabular} & \begin{tabular}[t]{@{}c@{}} 0.3262*** (0.0968) \\ \relax [0.0127] \end{tabular} & \begin{tabular}[t]{@{}c@{}} -0.3230*** (0.0937) \\ \relax [-0.0129] \end{tabular} \\
$is\_exclusive$ & \begin{tabular}[t]{@{}c@{}} 0.5348 (0.6418) \\ \relax [0.0186] \end{tabular} & \begin{tabular}[t]{@{}c@{}} 0.4576 (0.6761) \\ \relax [0.0167] \end{tabular} & -- & -- \\
$\ln s^{bal}$ & \begin{tabular}[t]{@{}c@{}} -0.1419*** (0.0397) \\ \relax [-0.0049] \end{tabular} & \begin{tabular}[t]{@{}c@{}} -0.2257*** (0.0377) \\ \relax [-0.0082] \end{tabular} & \begin{tabular}[t]{@{}c@{}} -0.1501*** (0.0381) \\ \relax [-0.0058] \end{tabular} & -- \\
$\ln t^{net}$ & \begin{tabular}[t]{@{}c@{}} -0.2156*** (0.0394) \\ \relax [-0.0075] \end{tabular} & -- & -- & \begin{tabular}[t]{@{}c@{}} -0.0039 (0.0433) \\ \relax [-0.0002] \end{tabular} \\
$\ln h$ & \begin{tabular}[t]{@{}c@{}} 2.4069*** (0.1725) \\ \relax [0.0836] \end{tabular} & \begin{tabular}[t]{@{}c@{}} 1.8489*** (0.1228) \\ \relax [0.0676] \end{tabular} & \begin{tabular}[t]{@{}c@{}} 1.2298*** (0.0952) \\ \relax [0.0479] \end{tabular} & \begin{tabular}[t]{@{}c@{}} 1.9997*** (0.1900) \\ \relax [0.0800] \end{tabular} \\
$\ln t^{bal}$ & \begin{tabular}[t]{@{}c@{}} -0.1047 (0.0969) \\ \relax [-0.0036] \end{tabular} & \begin{tabular}[t]{@{}c@{}} -0.0326 (0.0930) \\ \relax [-0.0012] \end{tabular} & \begin{tabular}[t]{@{}c@{}} 0.4650*** (0.0838) \\ \relax [0.0181] \end{tabular} & -- \\
\addlinespace
$\ln\text{Assets}_{firm}$ & \begin{tabular}[t]{@{}c@{}} 0.0068 (0.0295) \\ \relax [0.0002] \end{tabular} & \begin{tabular}[t]{@{}c@{}} -0.0186 (0.0273) \\ \relax [-0.0007] \end{tabular} & \begin{tabular}[t]{@{}c@{}} -0.0389 (0.0240) \\ \relax [-0.0015] \end{tabular} & \begin{tabular}[t]{@{}c@{}} -0.0143 (0.0250) \\ \relax [-0.0006] \end{tabular} \\
$\text{Lev}_{firm}$ & \begin{tabular}[t]{@{}c@{}} -0.0075 (0.0624) \\ \relax [-0.0003] \end{tabular} & \begin{tabular}[t]{@{}c@{}} 0.0790 (0.0554) \\ \relax [0.0029] \end{tabular} & \begin{tabular}[t]{@{}c@{}} -0.0729 (0.0539) \\ \relax [-0.0028] \end{tabular} & \begin{tabular}[t]{@{}c@{}} -0.0711 (0.0553) \\ \relax [-0.0028] \end{tabular} \\
$\text{ROA}_{firm}$ & \begin{tabular}[t]{@{}c@{}} -0.0089 (0.0152) \\ \relax [-0.0003] \end{tabular} & \begin{tabular}[t]{@{}c@{}} -0.0092 (0.0147) \\ \relax [-0.0003] \end{tabular} & \begin{tabular}[t]{@{}c@{}} -0.0019 (0.0151) \\ \relax [-0.0001] \end{tabular} & \begin{tabular}[t]{@{}c@{}} -0.0109 (0.0155) \\ \relax [-0.0004] \end{tabular} \\
$\text{Tang}$ & \begin{tabular}[t]{@{}c@{}} 0.1437 (0.4177) \\ \relax [0.0050] \end{tabular} & \begin{tabular}[t]{@{}c@{}} 0.0671 (0.3951) \\ \relax [0.0025] \end{tabular} & \begin{tabular}[t]{@{}c@{}} 0.0790 (0.3974) \\ \relax [0.0031] \end{tabular} & \begin{tabular}[t]{@{}c@{}} 0.2599 (0.3925) \\ \relax [0.0104] \end{tabular} \\
$\ln\text{Assets}_{bank}$ & \begin{tabular}[t]{@{}c@{}} -0.1340 (0.0957) \\ \relax [-0.0047] \end{tabular} & \begin{tabular}[t]{@{}c@{}} -0.2778*** (0.0893) \\ \relax [-0.0102] \end{tabular} & \begin{tabular}[t]{@{}c@{}} -0.5687*** (0.0916) \\ \relax [-0.0221] \end{tabular} & \begin{tabular}[t]{@{}c@{}} -0.3096*** (0.0854) \\ \relax [-0.0124] \end{tabular} \\
$\text{Lev}_{bank}$ & \begin{tabular}[t]{@{}c@{}} -0.0014 (0.0161) \\ \relax [-0.0000] \end{tabular} & \begin{tabular}[t]{@{}c@{}} -0.0164 (0.0170) \\ \relax [-0.0006] \end{tabular} & \begin{tabular}[t]{@{}c@{}} -0.0446*** (0.0163) \\ \relax [-0.0017] \end{tabular} & \begin{tabular}[t]{@{}c@{}} -0.0482*** (0.0146) \\ \relax [-0.0019] \end{tabular} \\
$\text{ROA}_{bank}$ & \begin{tabular}[t]{@{}c@{}} -0.0139 (0.1590) \\ \relax [-0.0005] \end{tabular} & \begin{tabular}[t]{@{}c@{}} -0.0751 (0.1781) \\ \relax [-0.0027] \end{tabular} & \begin{tabular}[t]{@{}c@{}} -0.2251 (0.1708) \\ \relax [-0.0088] \end{tabular} & \begin{tabular}[t]{@{}c@{}} -0.1340 (0.1697) \\ \relax [-0.0054] \end{tabular} \\
\hline
Observations & 9,379 & 9,379 & 9,379 & 9,379 \\
Pseudo $R^2$ & 0.380 & 0.364 & 0.316 & 0.303 \\
\hline
\hline
\multicolumn{5}{c}{\bf Panel B: Unconsolidated Sample} \\
\hline
\hline
Variable & 1. Empirical & 2. Emp. (No $s^{net}$, $t^{net}$) & 3. Null Net & 4. Null Bal \\ \hline
Intercept & \begin{tabular}[t]{@{}c@{}} -0.4963 \\ (1.0492) \end{tabular} & \begin{tabular}[t]{@{}c@{}} -0.4038 \\ (0.9894) \end{tabular} & \begin{tabular}[t]{@{}c@{}} -0.5513 \\ (0.9325) \end{tabular} & \begin{tabular}[t]{@{}c@{}} -0.4915 \\ (1.4633) \end{tabular} \\
$\ln s^{net}$ & \begin{tabular}[t]{@{}c@{}} -0.1909*** (0.0304) \\ \relax [-0.0065] \end{tabular} & -- & -- & \begin{tabular}[t]{@{}c@{}} -0.0843*** (0.0206) \\ \relax [-0.0035] \end{tabular} \\
$\ln k$ & \begin{tabular}[t]{@{}c@{}} 1.7576*** (0.1183) \\ \relax [0.0599] \end{tabular} & \begin{tabular}[t]{@{}c@{}} 1.4120*** (0.0990) \\ \relax [0.0523] \end{tabular} & \begin{tabular}[t]{@{}c@{}} 0.2106*** (0.0791) \\ \relax [0.0084] \end{tabular} & \begin{tabular}[t]{@{}c@{}} 0.1365 (0.0915) \\ \relax [0.0056] \end{tabular} \\
$is\_exclusive$ & \begin{tabular}[t]{@{}c@{}} 0.8567* (0.4386) \\ \relax [0.0292] \end{tabular} & \begin{tabular}[t]{@{}c@{}} 0.9766** (0.4032) \\ \relax [0.0362] \end{tabular} & -- & -- \\
$\ln s^{bal}$ & \begin{tabular}[t]{@{}c@{}} -0.1330*** (0.0195) \\ \relax [-0.0045] \end{tabular} & \begin{tabular}[t]{@{}c@{}} -0.1831*** (0.0188) \\ \relax [-0.0068] \end{tabular} & \begin{tabular}[t]{@{}c@{}} -0.1049*** (0.0179) \\ \relax [-0.0042] \end{tabular} & -- \\
$\ln t^{net}$ & \begin{tabular}[t]{@{}c@{}} -0.6575*** (0.0472) \\ \relax [-0.0224] \end{tabular} & -- & -- & \begin{tabular}[t]{@{}c@{}} -0.0176 (0.0377) \\ \relax [-0.0007] \end{tabular} \\
$\ln h$ & \begin{tabular}[t]{@{}c@{}} 2.8957*** (0.1485) \\ \relax [0.0987] \end{tabular} & \begin{tabular}[t]{@{}c@{}} 1.6822*** (0.1025) \\ \relax [0.0624] \end{tabular} & \begin{tabular}[t]{@{}c@{}} 1.0446*** (0.0733) \\ \relax [0.0418] \end{tabular} & \begin{tabular}[t]{@{}c@{}} 1.6657*** (0.1636) \\ \relax [0.0685] \end{tabular} \\
$\ln t^{bal}$ & \begin{tabular}[t]{@{}c@{}} -0.4542*** (0.0872) \\ \relax [-0.0155] \end{tabular} & \begin{tabular}[t]{@{}c@{}} -0.1280* (0.0746) \\ \relax [-0.0047] \end{tabular} & \begin{tabular}[t]{@{}c@{}} 0.3539*** (0.0761) \\ \relax [0.0142] \end{tabular} & -- \\
\addlinespace
$\ln\text{Assets}_{firm}$ & \begin{tabular}[t]{@{}c@{}} 0.0000 (0.0462) \\ \relax [0.0000] \end{tabular} & \begin{tabular}[t]{@{}c@{}} -0.1141*** (0.0426) \\ \relax [-0.0042] \end{tabular} & \begin{tabular}[t]{@{}c@{}} -0.1822*** (0.0408) \\ \relax [-0.0073] \end{tabular} & \begin{tabular}[t]{@{}c@{}} -0.2173*** (0.0476) \\ \relax [-0.0089] \end{tabular} \\
$\text{Lev}_{firm}$ & \begin{tabular}[t]{@{}c@{}} 0.0020 (0.0014) \\ \relax [0.0001] \end{tabular} & \begin{tabular}[t]{@{}c@{}} 0.0019 (0.0013) \\ \relax [0.0001] \end{tabular} & \begin{tabular}[t]{@{}c@{}} 0.0016 (0.0013) \\ \relax [0.0001] \end{tabular} & \begin{tabular}[t]{@{}c@{}} 0.0018 (0.0012) \\ \relax [0.0001] \end{tabular} \\
$\text{ROA}_{firm}$ & \begin{tabular}[t]{@{}c@{}} 0.0080 (0.0107) \\ \relax [0.0003] \end{tabular} & \begin{tabular}[t]{@{}c@{}} 0.0138 (0.0102) \\ \relax [0.0005] \end{tabular} & \begin{tabular}[t]{@{}c@{}} -0.0040 (0.0096) \\ \relax [-0.0002] \end{tabular} & \begin{tabular}[t]{@{}c@{}} -0.0076 (0.0093) \\ \relax [-0.0003] \end{tabular} \\
$\text{Tang}$ & \begin{tabular}[t]{@{}c@{}} -0.1886 (0.1836) \\ \relax [-0.0064] \end{tabular} & \begin{tabular}[t]{@{}c@{}} -0.1612 (0.1720) \\ \relax [-0.0060] \end{tabular} & \begin{tabular}[t]{@{}c@{}} -0.1756 (0.1613) \\ \relax [-0.0070] \end{tabular} & \begin{tabular}[t]{@{}c@{}} -0.0840 (0.1606) \\ \relax [-0.0035] \end{tabular} \\
$\ln\text{Assets}_{bank}$ & \begin{tabular}[t]{@{}c@{}} 0.3935*** (0.0938) \\ \relax [0.0134] \end{tabular} & \begin{tabular}[t]{@{}c@{}} -0.1687** (0.0812) \\ \relax [-0.0063] \end{tabular} & \begin{tabular}[t]{@{}c@{}} -0.3839*** (0.0958) \\ \relax [-0.0154] \end{tabular} & \begin{tabular}[t]{@{}c@{}} -0.1190 (0.1085) \\ \relax [-0.0049] \end{tabular} \\
$\text{Lev}_{bank}$ & \begin{tabular}[t]{@{}c@{}} 0.0042 (0.0135) \\ \relax [0.0001] \end{tabular} & \begin{tabular}[t]{@{}c@{}} 0.0186 (0.0124) \\ \relax [0.0007] \end{tabular} & \begin{tabular}[t]{@{}c@{}} -0.0132 (0.0133) \\ \relax [-0.0005] \end{tabular} & \begin{tabular}[t]{@{}c@{}} -0.0595*** (0.0134) \\ \relax [-0.0024] \end{tabular} \\
$\text{ROA}_{bank}$ & \begin{tabular}[t]{@{}c@{}} -0.2530 (0.1778) \\ \relax [-0.0086] \end{tabular} & \begin{tabular}[t]{@{}c@{}} 0.0055 (0.1274) \\ \relax [0.0002] \end{tabular} & \begin{tabular}[t]{@{}c@{}} 0.0563 (0.1056) \\ \relax [0.0023] \end{tabular} & \begin{tabular}[t]{@{}c@{}} -0.3488** (0.1438) \\ \relax [-0.0143] \end{tabular} \\
\hline
Observations & 11,200 & 11,200 & 11,200 & 11,200 \\
Pseudo $R^2$ & 0.396 & 0.340 & 0.275 & 0.253 \\
\hline \hline
\end{tabular*}
\caption{Robustness check for link formation: comparison of the predictive power of empirical vs. null model degrees for the full competition model (Model 3). 
Column `1. Empirical' is the baseline full model. 
Column `2. Emp. (No $S^{net}$)' excludes network strength. 
Column `3. Null Net' uses $k$ obtained from the \textit{Net} null. 
Column `4. Null Bal' uses $k$ obtained from the \textit{Bal} null. 
Coefficients are reported with standard errors in parentheses. Significance: *** $p<0.01$, ** $p<0.05$, * $p<0.1$. Values in square brackets $[\dots]$ represent Average Marginal Effects (AME).}
\label{tab:logit_robustness}
\end{table*}

\clearpage

\begin{table*}[!ht]
\centering
\renewcommand{\arraystretch}{1.1} % Riduce lo spazio tra le righe per far rientrare la tabella
\footnotesize
\begin{tabular*}{\textwidth}{@{\extracolsep{\fill}}lccccc}
\hline
\hline
\multicolumn{5}{c}{\bf Panel A: Consolidated Sample} \\
\hline
\hline
Variable & 1. Empirical & 2. Emp. (No $S^{net}$) & 3. Null Net & 4. Null Bal \\ \hline
Intercept & \begin{tabular}[t]{@{}c@{}} -5.3488*** (1.6236) \end{tabular} & \begin{tabular}[t]{@{}c@{}} -2.7824 (1.7996) \end{tabular} & \begin{tabular}[t]{@{}c@{}} 3.4965** (1.5774) \end{tabular} & \begin{tabular}[t]{@{}c@{}} -0.4055 (1.8539) \end{tabular} \\
$\ln s^{net}$ & \begin{tabular}[t]{@{}c@{}} 0.7147*** (0.0641) \end{tabular} & -- & -- & \begin{tabular}[t]{@{}c@{}} 0.0837** (0.0420) \end{tabular} \\
$\ln k$ & \begin{tabular}[t]{@{}c@{}} -1.0396*** (0.1510) \end{tabular} & \begin{tabular}[t]{@{}c@{}} -0.0584 (0.1646) \end{tabular} & \begin{tabular}[t]{@{}c@{}} 1.5412*** (0.1290) \end{tabular} & \begin{tabular}[t]{@{}c@{}} 0.8468*** (0.1338) \end{tabular} \\
$is\_exclusive$ & \begin{tabular}[t]{@{}c@{}} 6.9294*** (0.7535) \end{tabular} & \begin{tabular}[t]{@{}c@{}} 0.5090 (0.4707) \end{tabular} & -- & -- \\
$\ln s^{bal}$ & \begin{tabular}[t]{@{}c@{}} 0.0250 (0.0466) \end{tabular} & \begin{tabular}[t]{@{}c@{}} 0.1474** (0.0591) \end{tabular} & \begin{tabular}[t]{@{}c@{}} -0.0864** (0.0358) \end{tabular} & -- \\
$\ln t^{net}$ & \begin{tabular}[t]{@{}c@{}} 0.0701 (0.0496) \end{tabular} & -- & -- & \begin{tabular}[t]{@{}c@{}} 0.0809** (0.0402) \end{tabular} \\
$\ln h$ & \begin{tabular}[t]{@{}c@{}} -0.0930 (0.2099) \end{tabular} & \begin{tabular}[t]{@{}c@{}} 0.1017 (0.1730) \end{tabular} & \begin{tabular}[t]{@{}c@{}} 0.5167*** (0.1355) \end{tabular} & \begin{tabular}[t]{@{}c@{}} 0.0856 (0.2383) \end{tabular} \\
$\ln t^{bal}_{bank}$ & \begin{tabular}[t]{@{}c@{}} 0.0212 (0.1148) \end{tabular} & \begin{tabular}[t]{@{}c@{}} -0.1264 (0.1421) \end{tabular} & \begin{tabular}[t]{@{}c@{}} -0.0250 (0.0999) \end{tabular} & -- \\
\addlinespace
$\ln\text{Assets}_{firm}$ & \begin{tabular}[t]{@{}c@{}} 0.0370 (0.0419) \end{tabular} & \begin{tabular}[t]{@{}c@{}} 0.1377** (0.0596) \end{tabular} & \begin{tabular}[t]{@{}c@{}} 0.1017** (0.0398) \end{tabular} & \begin{tabular}[t]{@{}c@{}} 0.0599 (0.0509) \end{tabular} \\
$\text{Lev}_{firm}$ & \begin{tabular}[t]{@{}c@{}} -0.0081 (0.0905) \end{tabular} & \begin{tabular}[t]{@{}c@{}} 0.1151 (0.1044) \end{tabular} & \begin{tabular}[t]{@{}c@{}} 0.0524 (0.0862) \end{tabular} & \begin{tabular}[t]{@{}c@{}} 0.0513 (0.0958) \end{tabular} \\
$\text{ROA}_{firm}$ & \begin{tabular}[t]{@{}c@{}} -0.0024 (0.0153) \end{tabular} & \begin{tabular}[t]{@{}c@{}} -0.0320** (0.0157) \end{tabular} & \begin{tabular}[t]{@{}c@{}} -0.0233* (0.0132) \end{tabular} & \begin{tabular}[t]{@{}c@{}} 0.0011 (0.0152) \end{tabular} \\
$\text{Tang}$ & \begin{tabular}[t]{@{}c@{}} 0.4286 (0.4236) \end{tabular} & \begin{tabular}[t]{@{}c@{}} 1.2200*** (0.4345) \end{tabular} & \begin{tabular}[t]{@{}c@{}} 0.2284 (0.3967) \end{tabular} & \begin{tabular}[t]{@{}c@{}} 1.0873*** (0.4189) \end{tabular} \\
$\ln\text{Assets}_{bank}$ & \begin{tabular}[t]{@{}c@{}} 0.3536*** (0.1177) \end{tabular} & \begin{tabular}[t]{@{}c@{}} 0.5528*** (0.1422) \end{tabular} & \begin{tabular}[t]{@{}c@{}} 0.1248 (0.1314) \end{tabular} & \begin{tabular}[t]{@{}c@{}} 0.3071** (0.1284) \end{tabular} \\
$\text{Lev}_{bank}$ & \begin{tabular}[t]{@{}c@{}} 0.0335 (0.0239) \end{tabular} & \begin{tabular}[t]{@{}c@{}} 0.0194 (0.0261) \end{tabular} & \begin{tabular}[t]{@{}c@{}} 0.0197 (0.0238) \end{tabular} & \begin{tabular}[t]{@{}c@{}} 0.0154 (0.0256) \end{tabular} \\
$\text{ROA}_{bank}$ & \begin{tabular}[t]{@{}c@{}} -0.1411 (0.2599) \end{tabular} & \begin{tabular}[t]{@{}c@{}} 0.0776 (0.3231) \end{tabular} & \begin{tabular}[t]{@{}c@{}} -0.1614 (0.2691) \end{tabular} & \begin{tabular}[t]{@{}c@{}} -0.0109 (0.3036) \end{tabular} \\
\hline
Observations & 486 & 486 & 486 & 486 \\
$R^2$ & 0.491 & 0.298 & 0.514 & 0.385 \\
\hline
\hline
\multicolumn{5}{c}{\bf Panel B: Unconsolidated Sample} \\
\hline
\hline
Variable & 1. Empirical & 2. Emp. (No $S^{net}$) & 3. Null Net & 4. Null Bal \\ \hline
Intercept & \begin{tabular}[t]{@{}c@{}} -4.4380*** (1.4900) \end{tabular} & \begin{tabular}[t]{@{}c@{}} -4.7431*** (1.5341) \end{tabular} & \begin{tabular}[t]{@{}c@{}} 3.1834* (1.6334) \end{tabular} & \begin{tabular}[t]{@{}c@{}} -3.2843 (2.2420) \end{tabular} \\
$\ln s^{net}$ & \begin{tabular}[t]{@{}c@{}} 0.5350*** (0.0620) \end{tabular} & -- & -- & \begin{tabular}[t]{@{}c@{}} 0.0234 (0.0280) \end{tabular} \\
$\ln k$ & \begin{tabular}[t]{@{}c@{}} -0.5266*** (0.1496) \end{tabular} & \begin{tabular}[t]{@{}c@{}} 0.3487** (0.1577) \end{tabular} & \begin{tabular}[t]{@{}c@{}} 1.6159*** (0.1263) \end{tabular} & \begin{tabular}[t]{@{}c@{}} 0.9216*** (0.1760) \end{tabular} \\
$is\_exclusive$ & \begin{tabular}[t]{@{}c@{}} 5.0013*** (0.6711) \end{tabular} & \begin{tabular}[t]{@{}c@{}} 0.4244 (0.4206) \end{tabular} & -- & -- \\
$\ln s^{bal}$ & \begin{tabular}[t]{@{}c@{}} -0.0856*** (0.0213) \end{tabular} & \begin{tabular}[t]{@{}c@{}} -0.0794*** (0.0245) \end{tabular} & \begin{tabular}[t]{@{}c@{}} -0.1177*** (0.0147) \end{tabular} & -- \\
$\ln t^{net}$ & \begin{tabular}[t]{@{}c@{}} -0.0168 (0.0399) \end{tabular} & -- & -- & \begin{tabular}[t]{@{}c@{}} -0.0010 (0.0316) \end{tabular} \\
$\ln h$ & \begin{tabular}[t]{@{}c@{}} 0.1174 (0.1769) \end{tabular} & \begin{tabular}[t]{@{}c@{}} 0.1239 (0.1241) \end{tabular} & \begin{tabular}[t]{@{}c@{}} 0.2259* (0.1174) \end{tabular} & \begin{tabular}[t]{@{}c@{}} -0.4318** (0.1849) \end{tabular} \\
$\ln t^{bal}$ & \begin{tabular}[t]{@{}c@{}} -0.2714*** (0.0547) \end{tabular} & \begin{tabular}[t]{@{}c@{}} -0.3040*** (0.0592) \end{tabular} & \begin{tabular}[t]{@{}c@{}} -0.1864*** (0.0601) \end{tabular} & -- \\
\addlinespace
$\ln\text{Assets}_{firm}$ & \begin{tabular}[t]{@{}c@{}} 0.3374*** (0.0754) \end{tabular} & \begin{tabular}[t]{@{}c@{}} 0.6528*** (0.0659) \end{tabular} & \begin{tabular}[t]{@{}c@{}} 0.1792** (0.0714) \end{tabular} & \begin{tabular}[t]{@{}c@{}} 0.1811* (0.1095) \end{tabular} \\
$\text{Lev}_{firm}$ & \begin{tabular}[t]{@{}c@{}} 0.0011 (0.0015) \end{tabular} & \begin{tabular}[t]{@{}c@{}} 0.0031** (0.0014) \end{tabular} & \begin{tabular}[t]{@{}c@{}} 0.0018 (0.0012) \end{tabular} & \begin{tabular}[t]{@{}c@{}} 0.0006 (0.0013) \end{tabular} \\
$\text{ROA}_{firm}$ & \begin{tabular}[t]{@{}c@{}} 0.0090 (0.0141) \end{tabular} & \begin{tabular}[t]{@{}c@{}} 0.0207 (0.0160) \end{tabular} & \begin{tabular}[t]{@{}c@{}} 0.0322*** (0.0112) \end{tabular} & \begin{tabular}[t]{@{}c@{}} -0.0014 (0.0137) \end{tabular} \\
$\text{Tang}$ & \begin{tabular}[t]{@{}c@{}} 0.3309 (0.2343) \end{tabular} & \begin{tabular}[t]{@{}c@{}} 0.3336 (0.2437) \end{tabular} & \begin{tabular}[t]{@{}c@{}} 0.2226 (0.2027) \end{tabular} & \begin{tabular}[t]{@{}c@{}} 0.3865 (0.2368) \end{tabular} \\
$\ln\text{Assets}_{bank}$ & \begin{tabular}[t]{@{}c@{}} 0.5049*** (0.1103) \end{tabular} & \begin{tabular}[t]{@{}c@{}} 0.5416*** (0.1156) \end{tabular} & \begin{tabular}[t]{@{}c@{}} 0.2953** (0.1328) \end{tabular} & \begin{tabular}[t]{@{}c@{}} 0.5594*** (0.1187) \end{tabular} \\
$\text{Lev}_{bank}$ & \begin{tabular}[t]{@{}c@{}} 0.0298* (0.0161) \end{tabular} & \begin{tabular}[t]{@{}c@{}} 0.0295 (0.0187) \end{tabular} & \begin{tabular}[t]{@{}c@{}} 0.0311* (0.0161) \end{tabular} & \begin{tabular}[t]{@{}c@{}} 0.0209 (0.0184) \end{tabular} \\
$\text{ROA}_{bank}$ & \begin{tabular}[t]{@{}c@{}} -0.1692 (0.1980) \end{tabular} & \begin{tabular}[t]{@{}c@{}} -0.0117 (0.2298) \end{tabular} & \begin{tabular}[t]{@{}c@{}} -0.0270 (0.1795) \end{tabular} & \begin{tabular}[t]{@{}c@{}} 0.0111 (0.2951) \end{tabular} \\
\hline
Observations & 575 & 575 & 575 & 575 \\
$R^2$ & 0.442 & 0.311 & 0.554 & 0.342 \\
\hline \hline
\end{tabular*}
\caption{Robustness check for loan sizing: comparison of the predictive power of empirical vs. null model degrees for the full competition model (Model 3). 
Column `1. Empirical' is the baseline full model. 
Column `2. Emp. (No $S^{net}$)' excludes network strength. 
Column `3. Null Net' uses $k$ obtained from the \textit{Net} null. 
Column `4. Null Bal' uses $k$ obtained from the \textit{Bal} null. 
Coefficients are reported with standard errors in parentheses. Significance: *** $p<0.01$, ** $p<0.05$, * $p<0.1$.}
\label{tab:ecapm_regression}
\end{table*}

\end{document}